\documentclass[12pt,thmsa]{article}
\usepackage{amsfonts}

\usepackage{amsmath}

\input{tcilatex}

\begin{document}

\author{by Bert Schroer \and FU-Berlin, Institut f\"{u}r Theoretische Physik, \and %
Arnimallee 14, 14195 Berlin, Germany \and presently: CBPF Rio de Janeiro
\and email: schroer@cbpf.br}
\title{\textit{Particle Physics and} \textit{QFT at the Turn of the Century: }\\
\textit{Old principles with new concepts} \\
(an essay on local quantum physics)\\
revised and updated version.\\
Invited contribution to the Issue 2000 of JMP\\
{\small Dedicated to John Roberts on the occasion of his 60}$^{th}${\small \
birthday}}
\date{February 2000}
\maketitle

\begin{abstract}
The present state of QFT is analysed from a new viewpoint whose mathematical
basis is the modular theory of von Neumann algebras. Its physical
consequences suggest new ways of dealing with interactions, symmetries,
Hawking-Unruh thermal properties and possibly also extensions of the scheme
of renormalized perturbation theory. Interactions are incorporated by using
the fact that the S-matrix is a relative modular invariant of the
interacting- relative to the incoming- net of wedge algebras. This new point
of view allows many interesting comparisions with the standard quantization
approach to QFT and is shown to be firmly rooted in the history of QFT. Its
radical ``change of paradigm'' aspect becomes particularily visible in the
quantum measurement problem.

\end{abstract}

\section{Looking at the Past with Hindsight}

To a contemporary observer the area which half a century ago was very
appropriately called particle- or high-energy- physics with QFT being its
main theoretical tool, has gradually lost its homogeneous presentation and
appears presently somewhat fractured into several highly specialized regions
whose mutual relations are often lost. Despite analogies one would be very
hard-pressed to interpret e.g. the standard perturbative formulation
(especially of gauge theory), conformal field theory and massive factorizing
d=1+1 models as manifestations of the same physical principles. For this
reason\ the value of controllable low-dimensional models of QFT as a
theoretical laboratory to understand and explore the general principles of
Local Quantum Physics \cite{Haag} has remained opaque, despite the
considerable sophistication of their formalism which went into their
presentation. As no other previous theory in its long history, including
Relativity and Quantum Mechanics, QFT has resisted construction (apart from
some low-dimensional superrenormalizable models) to the degree that we do
not know whether those operators and their correlation functions which one
postulates and perturbatively ``approximates'' really exists in the presence
of 4-dim. nontrivial interactions\footnote{%
Despite numerous attempts to convert this problem into a small nuisance
which will be repaired at the future Plank length physics, the problem did
not go away. The problem of mathematical consistency of physical principles
cannot be solved by referring it to the next still unknown layer of physical
reality. }. The coexistence of such a curious state of affairs for almost 70
years with a set of perturbatively consistent rules and recipes of a
stunning predictive power is the most remarkable enigmatic heritage and a
gift of the 20$^{th}$ century particle physics to the 21$^{st}.$

We will have little to say about string theory which has separated from the
S-matrix aspects of QFT more than 3 decades years ago and still undergoes
rapid changes. The reason is that in addition to the absence of any tangible
contact with the nature of particle physics, string theory has failed to
compare its underlying principles with those of QFT or to formulate its own
principles. A theory which claims to transcend QFT without offering at the
same time new physical principles by which its underlying philosophy can be
secured against physical equivalence \cite{Weinberg} with field theoretic
principles is difficult to position. and here we will not even try.

A different and potentially more productive kind of dissatisfaction with the
present state of particle theory results from theories with impressive
predictive power but whose conceptual basis leaves too much to be desired in
order to be considered in the long run as completed and mature theories.
Here the very successful Bohr-Sommerfeld theory could serve as an example if
its incorporation into QM which showed its transitory character would not
have happened so swiftly. Potential contemporary candidates are electro-weak
theory and quantum chromodynamics. Most of their theoretical discoveries and
crucial theoretical developments occurred in the first 5 years starting in
the late 60$^{ies},$ although some of the important experimental
verifications came much later$.$ Compared with the speed of theoretical
progress during a good part of the 20$^{th}$ century, the time from the
middle 70$^{ies}$ up to now begins to appear more and more as a time of
stagnation. The fact that an increasing number of renown theoretical
particle physicist have uneasy feelings to accept the present gauge
theoretic models extended by the Higgs mechanism as a mature description
which constitute a closed chapter in particle physics, shows that this is
more than a overcritical interpretation on my part.

Experience with past crisis in particle theory (vis. the ultraviolet
divergency crisis of the 40$^{ies}$ solved by the renormalization theory of
the 50$^{ies})$ suggests to use a combination of conservative adherence to
physical principles and leave the revolutionary changes on the side of
concepts and mathematical formalism.

Most of the remedies which for the last two decades enjoy popularity (as
e.g. string theory and physics based on noncommutative geometry) were
revolutionary on the side of physical principles as well\footnote{%
A closer look reveals that they are in fact amazingly conservative the side
of formalism (e.g. the use of functional integral representations of the
Lagrangian quantization approach or ad hoc noncommutative modifications
thereof).}. As the history which led up to renormalization theory has shown,
it is easier to be revolutionary if one allows modifications of principles
(e.g. postulating an elementary length or fundamental cutoff, abandoning QFT
in favor of a pure S-matrix approach) than to maintain principles and limit
the changes to new concepts (physical reparametrization, changing the
canonical formalism for causal perturbations). It is indicative that even
when a change of principles became unavoidable, as in the case of relativity
and quantum theory, there was an intense conceptual struggle with the old
principles including the use of sophisticated Gedankenexperiments. It seems
that this intellectually demanding art has been lost in the second half of
the 20$^{th}$ century.

In the following I would like to expose some recent ideas which maintain a
strictly conservative attitude on the side of physical principles. So our
wanderlust to step into the ``blue yonder'' (to borrow a phrase used by
Feynman) will be controlled by the valuable compass of physical principles
underlying local quantum physics and not by the extension of existing
formalisms. The scheme which allows the most natural and clear formulation
for our purposes is nowadays referred to as algebraic QFT (AQFT) or local
quantum physics (LQP). Its impractical and non-constructive aspects (of
which it often stood accused) belongs to the past, and in the following we
will go a long way to demonstrate this. LQP as eenriched by modular theory
contains both of the two most successful aspects of past particle physics:
the formalism of local quantum physics but blended and controlled with the
Wigner particle concept and a new modular role of the S-matrix.  

Since I do not want to pose too many technical/mathematical barriers around
these new ideas, I use the more flexible essay style (``statements'' instead
of ``theorems''). I also assume that the reader is familiar with the
standard framework of QFT (\cite{Wei}).

For motivation I will first present some weak spots in the standard textbook
approach to QFT. Most of the presentations start with the canonical
formalism (Heisenberg-Pauli) or with the (Euclidean) path-integral formalism
(Feynman). Both of them are closely related quantization procedures. This
means that they are based on a classical parallelism starting from a
classical Lagrangian or Hamiltonian\footnote{%
In the case of Fermions it has been standard praxis (Berezin) to invent a
classical reality in form of Grassmann dynamical variables in order to
extend the quantization parallelism.} in analogy to the way quantum
mechanical systems are defined (and named after their classical
Hamiltonian). But there is one significant difference to the quantum
mechanical case. Whereas in the latter the canonical formalism and the
Feynman-Kac path-integral representation have a firm mathematical status
even in the presence of interactions, the use of these quantization methods
in QFT is (with the already mentioned exception of free fields and
superrenormalizable interactions) what one may call more of an ``artistic''
nature. This means that although the quantization requirements offer enough
guide to start perturbative calculations, the final renormalized answers do
not fulfil the original requirements: \textit{the renormalized physical
correlation functions simply do not obey the canonical commutation relations
nor are they Feynman-Kac representable!} The only generically remaining
structure unaffected by renormalization is Einstein causality/locality i.e.
the statement of mutual (anti)commutation of fields separated by spacelike
distances. In view of this delicate fact and despite the resulting lack of a
logical conceptual balance between the quantization requirement and the
physical renormalized answers, quantization in this sense became an accepted
fait accompli. The remarkable success swept aside worries for what appeared
just small mathematical imperfections. 

What enhanced the willingness of many physicists to live comfortably with
this conceptual flaw of the formulation of QFT was the fact that their
mathematical friends also became attracted to the differential-geometric
appeal of path integral quantization and often succumbed to its delicate
artistic fascination to such a degree that its conceptual and mathematical
flaws were ignored and the artistic computational tools became accepted as a
kind of experimental mathematics (and in several cases even received the
blessing of mathematicians). There is a lot of irony in the present state of
affairs where QM (for which the Feynman-Kac setting is rigorous but in
praxis too difficult and time-consuming) is presented with operator methods,
and on the other hand QFT (for which the method is a nice but artistic
device to get calculations started) is done almost exclusively in
path-integral formulation. Anybody who tried to give a physically balanced
course on QM using path integrals knows these problems.

There exists an alternative method of deforming free fields with interaction
Wick-polynomials within the setting of causal perturbation which uses the
above mentioned fact that causality (and not the Feynman-Kac
representability) survives renormalization. The interaction polynomials in
terms of free fields enter the causality- and unitarity- based equation for
time-ordered or retarded function as a perturbative input. All iterative
steps are then shown to be uniquely fixed by the mentioned principle and 
minimality requirements for an order-independent minimal scale dimension.
The mathematical problem is the extension of time-ordered distributions from
a certain subspace of test functions with nondiagonal support to such
containing supports with coalescing points. There is no actual infinity and
the difference renormalizable/nonrenormalizable is the implementability of
such a minimality requirement (which is tantamount to a unique theory with a
finite number of physical parameters). This method explains the infinities
of many of the textbook quantization method as a result of their unwarranted
relation to classical structures. In other words the prize of infinities of
the old classical particle models of Poincar\'{e} and Lorentz has entered
nolens volens via quantization and represent a technical nuisance \footnote{%
According to Wigner's analysis particles in QFT enter (to the degree that
the QFT possesses them) automatically via the representation theory of the
Poicar\'{e} group; there is no room for seperate particle models ala Poincar%
\'{e}/Lorentz.} which needs repair (as first pointed out by Kramers).
Despite differences in the conceptual setting the renormalized results of
all approaches (with or without intermediate infinities) are identical and
the apparent restrictive relation between the possible existence and
renormalizability of a theory and the ``good short distance behavior'' of
those particular ``field coordinatizations'' in terms of which the
interaction density was defined is common to all approaches which use
pointlike fields.

With this remark we have come to the point of departure of the new framework
from the old setting: \textit{the substitution of individual fields by nets
of algebras corresponding to spacetime regions.} This step is to be seen in
analogy to the transition of old fashioned geometry in terms of coordinates
to modern coordinate-free intrinsic differential geometry.

There were strong historical indications pointing towards a
field-coordinate-free formulation of local quantum physics\footnote{%
Since it is quite awkward to use the terminology ``QFT without pointlike
fields'', we follow Haag \cite{Haag}\ and use Local Quantum Physics or
algebraic QFT, in particular whenever we want the reader not to think in
terms of the standard textbook methods.}; one of the earliest was the
observation about the insensitivity of the (on-shell) scattering matrix with
respect to changes of the interpolating local fields. In the traditional
setting of Lagrangian this was done by carrying out extremely formal field
transformations. As in geometry one meets of course also preferred
field-coordinates which have characteristic intrinsic properties; notably
conserved Noether currents and other natural local objects which result from
the localization of (global) symmetries or have a direct relation to
superselected charges. I would even go as far as saying that it was
basically the arbitrary ad hoc nature of selection of particular fields in
particle physics which led to the (hard to understand from a contemporary
point of view) sometimes fanatical cleansing attitude against QFT (which
even entered the publications of some S-matrix purists). Our modular
localization approach will demonstrate, that also the opposite ideology
against S-matrix theory (for quotations of famous sayings see \cite{White})
is unwarranted. Since S is an important relative modular invariant, a
constructive method based on modular theory should use and construct it
together with the local algebras and not only at the end. Our approach
combines on- and off- shell aspects in one formalism and in particular
presents the construction of the observable algebras from an S-matrix point
of view without introducing individual fields; hence it accomplishes those
steps which in the old S-matrix theory were missing or even thought to be
impossible. 

In fact this coordinate-free formulation already exists for quite some time 
\cite{Haag}. Up to very recently it was limited to structural questions and
contributed little in the direction of classifications and investigations of
concrete models (a fact which perhaps also explains the widespread ignorance
about it). The main motivation for this essay is to inform the reader about
two new constructive ideas, both related to the Tomita-Takesaki modular
theory for wedge-localized algebras. The first idea uses ``polarization-
free generators'' of the wedge algebra whose structure is closely related to
the scattering matrix. This structure is e.g. behind the
bootstrap-formfactor program for d=1+1 factorizing theories. The second idea
is to relate a higher dimensional massive QFT to a finite number of
isomorphic copies of one chiral conformal field theory whose relative
positions in one Hilbert space are defined in terms of ``modular inclusions
and intersections''. In picturesque terms this should be thought of as some
sort of ``chiral scanning'' or AQFT-holography. One encodes the rather
complex structure of higher-dimensional massive QFT into a family of very
simple chiral conformal QFT and their relative modular position. Such
modular reformulations may also shed new light on the existence problem of
higher dimensional QFT since there is good control of existence of their
chiral conformal building blocks.

The organization of the content is as follows. As a ``warm up'' we explain
in the next section a presentation of interaction-free systems without the
use of field coordinates. We than use this formalism for a presentation of
the Hawking-Unruh thermal aspects of modular localization. The section
continues with a totally intrinsic characterization of what one means by
interactions. These results suggest to look at wedge algebras as the
smallest spacetime regions which offer the best compromise between particles
and fields; in fact if the often cited ''particle-field dualism'' makes any
sense at all, it is in this context of wedge localization. In the third
section we explain the relative modular invariance of the S-matrix. The
crucial concept here are certain wedge-localized operators which if applied
once to the vacuum even in the presence of interactions do not generate
particle/antiparticle vacuum polarization clouds but just pure one-particle
vectors. By specialization to 2-dim. models without real particle creation,
they are identified with Zamolodchikov-Faddeev operators which in this way
acquire for the first time a profound spacetime interpretation. We also
comment on wedge-localized states and operators in the presence of real pair
creation away from factorizing models. The section ends with a modular
extension of standard symmetries to ``hidden'' symmetries.

Section 3 presents the ``re-conquest'' of notions known from basic quantum
mechanics within LQP with the help of the ``split property''. In this
section the conceptual change of paradigm of the new approach becomes most
evident. In this section we also look at ``localization-entropy'', the other
thermal manifestation in addition to localization-temperature.

In the same futuristic last section I mention some potential areas of
applications where one expects the modular ideas to enlarge the conceptual
realm of models beyond what would be called ordinarily
``nonrenormalizable''. We also present various other poorely studied
consequences of modular theory, including an LQP version of ``holography''
and ``chiral conformal scanning''.

This essay is intended to fill some of the space left between two other
major articles on the present state of Local Quantum physics in this same
JMP\ issue; one is a broadly-based paper with a strong emphasis on recalling
the history and the spirit of times of particle physics during a good part
of the 20$^{th}$ century \cite{B-H}, and the other \cite{Bor} presents an
exhaustive account of more recent developments about modular stuctures of
LQP.

\section{Modular Structure of LQP}

For pedagogical explanations of the new modular concepts, the interaction
free theories are still the simplest. As in some of the textbooks (Haag,
Weinberg), one starts from the Wigner approach which assigns a unique
irreducible representation of the Poincar\'{e} group with every admissable
value of the mass and spin/helicity (m,s). The Wigner theory also preempts
the statistics of particles and assigns in the case of d=3+1, where the
particles can only be Fermions/Bosons (with the exception of the essentially
unexplored case of continuous spin), unique momentum space creation and
annihilation operators acting in a multiparticle Fock space. The uniqueness
is only lost at the moment one uses a manifestly local formalism in terms of
pointlike fields. The covariant field construction is synonymous with the
introduction of intertwiners between the unique Wigner $(m,s)$
representation and Lorentz-covariant momentum dependent spinorial (dotted
and undotted) tensors which under the homogenous L-group transform with the
irreducible $D^{\left[ A,B\right] }(\Lambda )$ matrices.

\begin{equation}
u(p)D^{(s)}(R(\Lambda ,p))=D^{[A,B]}(\Lambda )u(\Lambda ^{-1}p)  \label{1}
\end{equation}
The only restriction is: 
\begin{equation}
\mid A-B\mid \leq s\leq A+B  \label{2}
\end{equation}
which leaves infinitely many $A,B$ (half integer) choices for a given $s$.
Here the $u(p)$ intertwiner is a rectangular matrix consisting of $2s+1$
column vectors $u(p,s_{3}),s_{3}=-s,...,+s$ of length $(2A+1)(2B+1)$. Its
explicit construction using Clebsch-Gordan methods can be found in
Weinberg's book \cite{Wei}. Analogously there exist antiparticle (opposite
charge) $v(p)$ intertwiners: $D^{(s)\ast }(R(\Lambda ,p)\longrightarrow
D^{[A,B]}(\Lambda )$. The covariant field is then of the form: 
\begin{eqnarray}
\psi ^{\lbrack A,B]}(x) &=&\frac{1}{(2\pi )^{3/2}}\int
(e^{-ipx}\sum_{s_{3}}u(p_{1},s_{3})a(p_{1},s_{3})+  \label{field} \\
&&+e^{ipx}\sum_{s_{s}}v(p_{1},s_{3})b^{\ast }(p_{1},s_{3}))\frac{d^{3}p}{%
2\omega }  \notag
\end{eqnarray}

Since the range of the A and B (undotted/dotted) spinors is arbitrary apart
from the fact that they must fulfil the inequality with respect to the given
physical spin s\footnote{%
For the massless case the helicity inequalities with respect to the
spinorial indices are more restrictive, but one Wigner representation has
still a countably infinite number of covariant representations.}, the number
of covariant fields is countably infinite. Fortunately it turns out that
this loss of uniqueness does not cause any harm in particle physics. If one
defines the algebras $\mathcal{P}(\mathcal{O})$ as the operator algebras
generated from the smeared field with supp $f\in \mathcal{O}$ \cite{Wightman}%
,one realises that these localized algebras do not depend on the
representative field chosen from the (m,s) class. In fact all the different
covariant fields which share the same creation/annihilation operators. This
gave rise to the Borchers equivalence classes of relatively local fields 
\cite{Wightman} which generalized the family of Wick p\"{o}lynomials to the
realm of interactions and gave a structural explanation of the insensitivity
of the S-operator.

\subsection{Modular Aspects of Wigner Particle Theory}

The conceptually and mathematically natural way to implement the idea of
independence of physics from different field coordinatizations is to use
instead of smeared unbounded fields (with their technically difficult domain
properties) the associated von Neumann algebras of bounded operators \cite
{Haag} which have lost there reference to particular field coordinates. In
the case at hand of the Wigner particle theory of free particles this step
recovers the Wigner uniqueness of (m,s) particle representations (which got
lost as a result of the introduction of covariant fields). The obvious
question is therefore: \textit{is it possible to extract the spacetime
indexed net of algebras directly from the Wigner theory without the
intermediate appearance of fields?} A question like this was probably on
Wigner's mind when he was looking (without success) for a relativistic
localization concept within his representation-theoretical framework.

Recently this question of covariant localizaton received a positive answer
as a result of the introduction of ``modular localization'' \cite{Sch1}\cite
{BGL}. The idea can be traced back to a seminal paper of Bisognano and
Wichmann in which it was shown that the modular Tomita-Takesaki theory \cite
{Haag} of von Neumann algebras has not only some deep use in quantum
statistical mechanics (as was already known from the Haag-Hugenholtz-Winnink
work which appeared at the same time as Tomita's notes \cite{Haag}) but is
also an inexorable part of field theoretic wedge-localization\footnote{%
It is important to emphasize that physicists have a significant share in the
discovery of modular theory in particular with physicists whose only contact
with this theory arose through ``non-commutative geometry'' without
revealing the natural physical origin.}. What one needs here is in some
sense the inverse of the Bisognano-Wichmann arguments namely the use of
modular theory for the actual construction of a net of wedge algebras and
their smaller descendents from intersections. Its adaptation to the case at
hand would look for a kind of pre-Tomita theory which can be formulated
within the Wigner theory and with the help of CCR/CAR functors preempts the
net structure of the interaction-free LQP. This is indeed feasible and the
resulting formalism is mathematically not more complicated than the
formalism of free fields. Since it has appeared in different publications, a
short description should suffice for the purpose of this essay.

The pre-modular theory alluded to is a generalization of the Tomita theory
to real Hilbert spaces positioned within a complex Hilbert space. For its
adaptation to the Wigner theory one starts with the boost transformation
associated with a wedge and its reflection transformation along the rim of
the wedge. For the standard x-t wedge these are the $\Lambda _{x-t}(\chi )$
Lorentz boost and the x-t reflection $r_{x-t}:$ ($x,t)\rightarrow (-x$,$-t)$
which according to well-known theorems is represented antiunitarily in the
Wigner theory\footnote{%
In case of charged particles the Wigner theory should be suitably extended
by a particle/antiparticle doubling.}. One then starts from the unitary
boost group $u(\Lambda (\chi )$ and forms by the standard functional
calculus the unbounded ``analytic continuation''. In terms of modular
notation we define 
\begin{eqnarray}
\frak{s} &=&\frak{j}\delta ^{\frac{1}{2}} \\
\frak{j} &=&U(r)  \notag \\
\delta ^{it} &=&u(\Lambda (-2\pi t))  \notag
\end{eqnarray}
where $u(\Lambda (\chi )$ and $u(r)$ are the unitary/antiunitary
representations of these geometric transformations in the (if necessary
doubled) Wigner theory. Note that U(r) is apart from a $\pi $-rotation
around the x-axis the one-particle version of the TCP operator. On the other
hand $\frak{s}$ is a very unusual object namely an unbounded antilinear
operator which on its domain is involutive $\frak{s}^{2}=1.$ The real
subspace 
\begin{equation}
\frak{s}\psi =\psi
\end{equation}
which consists of momentum space wave functions which are boundary values of
analytic functions in the lower $i\pi -$strip of the rapidity variable $%
\theta .$ The -1 eigenvalues of S do not give rise to a new problem since
multiplication of the +1 eigenfunctions with i convert them into the -1
eigenfunctions. The real subspace $H_{R}(W)$ is closed in the complex
Hilbert space topology but the complexification $H_{R}(W)$ gives a space
which is only dense in the complex Wigner space. This surprising fact (which
is the Wigner one-particle analogue of the Reeh-Schlieder denseness of local
field states in full quantum field theory) has no parallel in any other area
of quantum physics. It suggests that the above mentioned unusual property of
the S-operator may be the vehicle by which geometric physical properties of
space time localization are encoded into the abstract domain properties of
unbounded operators. Some rather straightforward checks reveal that this
interpretation is consistent: in the present setting this localization
interpretation gives consistency with the net properties of the spaces $%
H_{R}(\mathcal{O})$'s 
\begin{equation}
H_{R}(\mathcal{O})\equiv \cap _{W\supset \mathcal{O}}H_{R}(W)
\end{equation}
as well as with the conventional field theoretic construction using
pointlike fields where it agrees with localized covariant functions defined
in terms of support properties of Cauchy initial data. The relation of
Wigner subspaces and localized subalgebras is accomplished with the help of
the CCR or CAR functors which map real subspaces $H_{R}(\mathcal{O})$ into
von Neumann $\mathcal{A}(H_{R}(\mathcal{O}))$ subalgebras and which define a
limited but rigorous meaning of the word ``quantization'' 
\begin{equation}
J,\Delta ,S\frak{=\Gamma (j,\delta ,s)}
\end{equation}
where the functorial map $\Gamma $ carries the functions of the Wigner
theory into the Weyl operators in Fock space (for the fermionic CAR-algebras
there is an additional modification). Whereas the ``pre-modular'' operators
denoted by small letters act on the Wigner space, the modular operators $%
J,\Delta $ have an $Ad$ action on the von Neumann algebras which are
functorially related to the subspaces and which makes them objects of the
Tomita-Takesaki modular theory 
\begin{eqnarray}
&&SA\Omega =A^{\ast }\Omega ,\,S=J\Delta ^{\frac{1}{2}} \\
&&Ad\Delta ^{it}\mathcal{A}=\mathcal{A} \\
&&AdJ\mathcal{A}=\mathcal{A}^{\prime }  \notag
\end{eqnarray}
This time the S-operator is that of Tomita i.e. the unbounded densely
defined operator which relates the dense set $A\Omega $ to the dense set $%
A^{\ast }\Omega $ and gives $J$ and $\Delta ^{\frac{1}{2}}$ by polar
decomposition. The nontrivial miraculous properties of this decomposition
are the existence of an automorphism $\sigma _{\omega }(t)=Ad\Delta ^{it}$
which propagates operators within $\mathcal{A}$ and only depends on the
state $\omega $ (and not on the implementing vector $\Omega )$ and a that of
an antiunitary involution $J$ which maps $\mathcal{A}$ onto its commutant $%
\mathcal{A}^{\prime }.$ The theory of Tomita assures that these objects
exist in general if only $\Omega $ is a cyclic and separating vector with
respect to $\mathcal{A}.$ Our special case at hand, in which the algebras
and the modular objects are constructed functorially from the Wigner theory,
suggest that the modular structure for wedge algebras may always have a
geometrical significance with a fundamental physical interpretation in any
QFT. This is indeed true, and within the Wightman framework this was
established by Bisognano-Wichmann \cite{Haag}.

The existence of this coordinate-free formulation for interaction free
theories has immediate consequences. Although in the present form it is not
yet suited to incorporate interactions without the use of field coordinates,
it does shed an additional helpful light on the standard causal perturbation
theory. Among other things it formally explains why an interaction which has
been defined in terms of concrete free fields can be rewritten without
change of content in terms of any other field coordinates and that moreover
Euler-Lagrange coordinates which associate free fields with a bilinear zero
order Lagrangians $\mathcal{L\,}_{0}$ are not necessary in a real time
operator formulation.. Of course since an Euler-Lagrangian field
coordinatization exists for each free theory and physical results remain the
same if on properly tranforms the interaction density, the use of such field
is not a restriction of generality.

\subsection{Thermal Aspects of Modular Localization}

Another valuable suggestion which can be abstracted from the pre-modular
structure of the Wigner theory concerns thermal aspects which originate from
localization. In modular theory the dense set of vectors which are obtained
by applying (local) von Neumann algebras in standard position to the
standard (vacuum) vector forms a core for the Tomita operator $S.$ The
domain of $S$ can then be described in terms of the +1 (or -1) closed real
subspace of $S.$ In terms of the ``pre-modular'' objects $\frak{s}$ in
Wigner space and the modular Tomita operators $S$ in Fock space we introduce
the following nets of wedge-localized dense subspaces: 
\begin{equation}
H_{R}(W)+iH_{R}(W)=dom(s)\subset H_{Wigner}
\end{equation}

\begin{equation}
\mathcal{H}_{R}(W)+i\mathcal{H}_{R}(W)=dom(S)\subset \mathcal{H}_{Fock}
\end{equation}

These dense subspaces become Hilbert spaces in their own right if we use the
graph norm (the thermal norm) of the Tomita operators. For the $s$-operators
in Wigner space we have: 
\begin{eqnarray}
\left( f,g\right) _{Wigner} &\rightarrow &\left( f,g\right) _{G}\equiv
\left( f,g\right) _{Wig}+\overline{\left( sf,sg\right) }_{Wig}  \label{graph}
\\
&=&\left( f,g\right) _{Wig}+\left( f,\delta g\right) _{Wig}  \notag
\end{eqnarray}
This graph topology insures that the wave functions are strip-analytic in
the wedge rapidity $\theta $: 
\begin{eqnarray}
p_{0} &=&m(p_{\perp })\cosh \theta ,\,\,\,\,p_{1}=m(p_{\perp })\sinh \theta
,\,\,\,m(p_{\perp })=\sqrt{m^{2}+p_{\perp }^{2}}  \label{rap} \\
&&strip:0<Imz<\pi ,\,\,\,\,\,\,z=\theta _{1}+i\theta _{2}  \notag
\end{eqnarray}
where the ''G-finiteness'' (\ref{graph}) is precisely the analyticity
prerequisite for the validity of the KMS property for the two-point
function. In this way one finally arrives at (for scalar Bosons):

\begin{equation}
\left( f,g\right) _{Wig}^{W}\equiv \left( f,g\right) _{K,T=2\pi }
\end{equation}
where on the left hand side the Wigner inner product is restricted to $%
H_{R}(W)+iH_{R}(W)$ and the right hand side is the thermal inner product
which contains the characteristic thermal $\frac{\delta }{1-\delta }$ factor
where $\delta =e^{2\pi K}$ with $K\footnote{%
The Unruh Hamiltonian is different from the boost $K$ by a factor $\frac{1}{a%
}$ where $a$ is the Unruh acceleration.}$ the infinitesimal generator of the
L-boost. The fact that the boost $K\,$with a two-sided spectrum appears
instead of the one-sided bounded Hamiltonian $H$ reveals one difference
between the two situations. More explicitly for the heat bath temperature of
a Hamiltonian dynamics the modular operator $\delta =e^{-2\beta \mathbf{H}}$
is bounded on one particle wave functions, whereas the unboundedness of $%
\delta =e^{2\pi K}$ enforces the localization (strip analyticity) of the
Wigner wave functions i.e. the two-sidedness of the spectrum does not permit
a KMS state on the full algebra. $\,$In fact localization-temperatures are
inexorably linked with unbounded modular symmetry operators. With the
localization-temperature $T=2\pi $ in this way having been made manifest,
the only difference between localization-temperatures and heat bath
temperatures (for a system enclosed in a box described by a Gibbs formula)
on the level of field algebras in Fock space corresponds to the difference
between hyperfinite type $III_{1}$ and type $I$ von Neumann algebras. But
even this distinction disappears if one passes from the Gibbs box situation
to the infinite volume thermodynamic limit: the GNS reconstruction using the
limiting correlation functions reveals that the algebra has become
hyperfinite type III$_{1}.$

Passing from Wigner one-particle theory to free field theory we may now
consider matrix elements of wedge-localized operators between
wedge-localized multiparticle states. Then the KMS property allows to move
the wedge-localized particle state as an antiparticle with the analytically
continued rapidity $\theta +i\pi $ from the ket to the bra. The simplest
illustration is the two-particle matrix element of a free current of a
charged scalar field $j_{\mu }(x)=:\phi ^{\ast }\overset{\leftrightarrow }{%
\partial }_{\mu }\phi :$ . The analytic relation 
\begin{equation}
\left\langle p^{\prime }\left| j_{\mu }(0)\right| p\right\rangle =\underset{%
z\rightarrow \theta +i\pi }{anal.cont.}\left\langle 0\left| j_{\mu
}(0)\right| p,\bar{p}^{\prime }(z)\right\rangle  \label{cross}
\end{equation}
where $\bar{p}^{\prime }(z)$ represents the analytic rapidity
parametrization of the antiparticle is the simplest form of a crossing
relation. It is an identity which is known to hold also in each perturbative
order of renormalizable interacting theories and which together with
TCP-symmetry constitutes the most profound property of QFT. But it has never
been derived in sufficient generality within a nonperturbative framework of
QFT nor (different from TCP) has its relation to the causality and positive
energy property of QFT been adequately understood. It is often thought of as
a kind of on shell momentum space substitute for Einstein causality and
locality and its strengthened form, called Haag duality.

If crossing symmetry is really a general property of local QFT, then it
should be the on shell manifestation of the off shell KMS property for
modular wedge localization not only in the previous free case but also in
the presence of interactions. In fact we will show in the next section that
the main step towards a deeper understanding of crossing symmetry is the
existence of certain on-shell operators $\int F(x)f(x)dx$ ($suppf\in W)$
which generate the wedge algebra and upon application to the vacuum create a
one-particle state vector without the vacuum polarization clouds which are
characteristc for interacting operators in smaller than wedge lovalization
regions. We will call them polarization-free generators or PFG's. In the
case of d=1+1 factorizing models their mass shell Fouriertransforms satisfy
the Zamolodchikov-Faddeev algebraic relations\footnote{%
As will become clear in the next section, although these operators are
nonlocal, they generate the wedge localized algebra, and as a consequence
the modular KMS formalism is applicable to them.} in the momentum space
rapidity \cite{Za}, and the derivation of crossing symmetry is similar
(albeit more involved) to the previously mentioned case of formfactors in
free theories.

\subsection{Wedge Localization for Special Interactions}

A major challenge to ones conceptual abilities is the transposition of these
modular attempts to the realm of interactions. Here the first step should be
a clear intrinsic definition of what one means by interactions without the
use of e.g. Lagrangians, Feynman rules or other ways of computing but solely
based on intrinsic properties of correlation functions or nets of local
algebras . The example of Wick polynomials in the free Borchers class, which
despite their complicated looking vacuum correlation functions still
represent only free theories in the veil of different field coordinates,
gives a first taste of the magnitude of the problem. This will be addressed
in the next subsection.

We start with the Fock space of free massive Bosons or Fermions. In order to
save notation we will explain the main ideas first in the context of
selfconjugate (neutral) scalar Bosons. Using the Bose statistics we will use
for our definitions the ``natural'' rapidity-ordered notation for n-particle
state vectors 
\begin{equation}
a^{\ast }(\theta _{1})a^{\ast }(\theta _{2})...a^{\ast }(\theta _{n})\Omega
,\,\,\,\theta _{1}>\theta _{2}>...>\theta _{n}
\end{equation}
and define new creation operators $Z^{\ast }(\theta )$ in case of $\theta
_{i}>\theta >\theta _{i+1}$ and with the previous convention 
\begin{eqnarray}
&&Z^{\ast }(\theta )a^{\ast }(\theta _{1})...a^{\ast }(\theta
_{i})...a^{\ast }(\theta _{n})\Omega = \\
&&S(\theta -\theta _{1})...S(\theta -\theta _{i})a^{\ast }(\theta
_{1})...a^{\ast }(\theta _{i})a^{\ast }(\theta )...a^{\ast }(\theta
_{n})\Omega  \notag
\end{eqnarray}
With $Z(\theta )$ as the formal adjoint one finds the following two-particle
commutation relations 
\begin{eqnarray}
Z^{\ast }(\theta )Z^{\ast }(\theta ^{\prime }) &=&S(\theta -\theta ^{\prime
})Z^{\ast }(\theta ^{\prime })Z^{\ast }(\theta )  \label{ab} \\
Z(\theta )Z^{\ast }(\theta ^{\prime }) &=&S(\theta ^{\prime }-\theta
)Z^{\ast }(\theta ^{\prime })Z(\theta )+\delta (\theta -\theta ^{\prime }) 
\notag
\end{eqnarray}
where the formal $Z$ adjoint of $Z^{\ast }$ is defined in the standard way.
The $\ast -$algebra property requires $S(\theta )=S(\theta )^{\ast
}=S(\theta )^{-1}=S(-\theta ).$ Although our notation already preempted the
relation with the Zamolodchikov-Faddeev algebra, \ the conceptual setting
here is quite different. We do not demand that the structure function S is
the crossing symmetric S-matrix where certain poles represent bound states
of particles. Rather we will show that all these properties including their
physical interpretation are consequences of modular wedge localization of
PFG's formed from the $Z^{\prime }s.$ This structure leads in particular to 
\begin{eqnarray}
Z^{\ast }(\theta _{1})...Z^{\ast }(\theta _{n})\Omega &=&a^{\ast }(\theta
_{1})...a^{\ast }(\theta _{n})\Omega \\
Z^{\ast }(\theta _{n})...Z^{\ast }(\theta _{1})\Omega &=&\prod_{i>j}S(\theta
_{i}-\theta _{j})a^{\ast }(\theta _{1})...a^{\ast }(\theta _{n})\Omega 
\notag
\end{eqnarray}
for the natural/opposite order with all other cases between these extreme
orders. Note that for momentum space rapidities it is not necessary to say
something about coinciding rapidities since only the $L^{2}$
measure-theoretical sense and no continuity is relevant here. In fact the
mathematical control of these operators i.e. the norm inequalities involving
the number operator hold as for the standard creation/annihilation
operators. Let us now imitate the free field construction and ask about the
localization properties of these F-fields 
\begin{equation}
F(x)=\frac{1}{\sqrt{2\pi }}\int (e^{-ipx}Z(\theta )+h.c.)
\end{equation}
This field has all the standard properties of operator-valued tempered
distributions, but it cannot be local if $S$ depends on $\theta $ since the
on-shell property together with locality leads to the free field formula. In
fact it will turn out (see next section) that the smeared operators $%
F(f)=\int F(x)f(x)d^{2}x$ with 
\begin{equation}
suppf\in W_{0}=\left\{ x;\,\,x^{1}>\left| x^{0}\right| \right\}
\end{equation}
have their localization in the standard wedge $W$ and that, contrary to
smeared pointlike localized fields, the wedge localization cannot be
improved by improvements of the test function support inside $W.$ Instead
the only way to come to a local net of algebras (and, if needed, to their
pointlike field generators) is by intersecting oppositely localized wedge
algebras (see below). Anticipating their wedge localization properties these
operators are our first examples of \textbf{p}olarization \textbf{f}ree 
\textbf{g}enerators (PFG). Like free fields their one-time application to
the vacuum creates a one-particle state without a (vacuum) polarization
cloud admixture.

We want to show that the operators $F(f)$ are generators of a wedge
localized algebra 
\begin{equation}
\mathcal{A}(W)=alg\left\{ F(f);suppf\in W\right\}
\end{equation}
As in the case of free fields the algebra may be defined as the weak closure
of the C$^{\ast }$- algebra generated by the spectral projection operators
in the spectral resolution 
\begin{equation}
F(f)=\int \lambda dE_{f}(\lambda )
\end{equation}
We first show that n-point functions of the $F(f)^{\prime }s$ obey a KMS
condition with respect to the Lorentz-boost subgroup which leaves the wedge $%
W_{0}$ invariant if and only if the commutation functions (in addition to
their holomorphy properties in the $\theta $-strip) are crossing symmetric
which is the symmetry of reflections through the point $\theta =i\frac{\pi }{%
2}$ (with the additional appearance of the charge conjugation for
non-neutral particles). One can show the following statement

\textit{Statement}:(\cite{Sch1}) The KMS-thermal property of the wedge
algebra generated by the PFG's is equivalent to the crossing symmetry of the
S-matrix

\begin{eqnarray}
\left( \Omega ,F(f_{1^{^{\prime }}})F(f_{2^{^{\prime
}}})F(f_{2})F(f_{1})\Omega \right) &\equiv &\left\langle F(f_{1^{^{\prime
}}})F(f_{2^{^{\prime }}})F(f_{2})F(f_{1})\right\rangle _{therm} \\
&&\overset{KMS}{=}\left\langle F(f_{2^{^{\prime
}}})F(f_{2})F(f_{1})F(f_{1^{^{\prime }}}^{-2\pi i})\right\rangle _{therm} 
\notag \\
&\Leftrightarrow &S(\theta )=S(i\pi -\theta )
\end{eqnarray}
Here we only used the cyclic KMS property (the second line containing the
imaginary $2\pi $-shift) for the four-point function. The relation is
established by Fouriertransformation and contour shift $\theta \rightarrow
\theta -i\pi .$ One computes

\begin{eqnarray}
&&F(\hat{f}_{2})F(\hat{f}_{1})\Omega =\int \int f_{2}(\theta _{2}-i\pi
)f_{1}(\theta _{1}-i\pi )Z^{\ast }(\theta _{1})Z^{\ast }(\theta _{2})\Omega
+c.t. \\
&=&\int \int f_{2}(\theta _{2}-i\pi )f_{1}(\theta _{1}-i\pi )\{\chi
_{12}a^{\ast }(\theta _{1})a^{\ast }(\theta _{2})\Omega +  \notag \\
&&+\chi _{21}S(\theta _{2}-\theta _{1})a^{\ast }(\theta _{2})a^{\ast
}(\theta _{1})\Omega \}+c\Omega
\end{eqnarray}
where the $\chi $ are the characteristic function for the differently
permuted $\theta $-orders. The analogous formula for the bra-vector is used
to define the four-point function as an inner product. If S has a crossing
symmetric pole in the in the physical strip of S the contour shift will
produce an unwanted terms which wrecks the KMS relation. The only way out is
to modify the previous relation 
\begin{eqnarray}
F(\hat{f}_{2})F(\hat{f}_{1})\Omega &=&(F(\hat{f}_{2})F(\hat{f}_{1})\Omega
)_{scat}  \label{2F} \\
&&+\int d\theta f_{1}(\theta _{1}+i\theta _{b})f_{2}(\theta _{2}-i\theta
_{b})\left| \theta ,b\right\rangle \left\langle \theta ,b\left| Z^{\ast
}(\theta -i\theta _{b})Z^{\ast }(\theta +i\theta _{b})\right| \Omega
\right\rangle
\end{eqnarray}
The second contribution is compensated by the pole contribution from the
contour shift. In general the shift will produce an uncompensated term from
a crossed pole whose position is obtained by reflecting in the imaginary
axis around $i\frac{\pi }{2}.$ which creates the analogous crossed bound
state contribution. In our simplified selfconjugate model it is the same
term as above. In the presence of one or several poles one has to look at
higher point functions. Despite the different conceptual setting one obtains
the same formulas as those for the S-matrix bootstrap of factorizing models
and hence one is entitled to make use of the bootstrap technology in this
modular program. What is behind is the so-called GNS construction which
converts the numerical poles in S and its higher bound versions into new
states i.e. the original Fock-space structure has to be enlarged if we
initially forgot to include the b-particles. Even though the description of
the wedge algebra appears like QM, there is one important difference which
is worthwhile noticing. This is the principle of ``nuclear democracy''
between particles. In QM there is a hierarchy between fundamental and bound:
elementary states do not reappear as boundstates of others and in particular
not of composites of themselves. We will see in the following that this
realization of nuclear democracy for double cone algebras is not any more
with particles and their binding, but rather with charges and their fusion.
The reason is of course the appearance of polarization clouds below wedge
localization. This nuclear democracy idea was the basis of the S-matrix
bootstrap approach and was first made to work in special two-dimensional
situations in \cite{STW}\cite{BKTW}\cite{Zam}.

With the derivation of crossing symmetry and the bound state and fusion
structure of $S$ we achieved our aim to present an example of the
constructive power of the modular localization method. In fact our fusion
formulas for multi F-vectors correctly interpret the Z-formulas in \cite{Za} 
\cite{Fring} which if taken literally are not true. As an unexpected
gratification we also obtained the equivalence between the crossing symmetry
of particle physics and the thermal KMS properties of the Hawking-Unruh
effect.

Strictly speaking the check of the KMS property with the Lorentz-boost as
the automorphism of the wedge algebra does not yet prove that the modular
theory is completely geometric. If we could show that the Tomita involution
is equal to the TCP operator, we would be done. For this to hold, we define 
\begin{equation}
J=\mathbf{S}_{s}J_{0}
\end{equation}
This relation between the incoming Tomita involution $J_{0}$ and that of the
interacting theory $J$ is nothing else as the TCP transformation for the
scattering matrix in a general local QFT.

with $J_{0}$ being the Tomita involution (=TCP) for the wedge algebra of the
free field theory We can now directly check 
\begin{eqnarray}
&&\hat{Z}^{\ast }(\theta ):=JZ^{\ast }(\theta )J  \notag \\
&&\left[ \hat{Z}^{\ast }(\theta ),Z^{\ast }(\theta ^{\prime })\right] =0 \\
&&\left[ \hat{Z}(\theta ),Z^{\ast }(\theta ^{\prime })\right] =\delta
(\theta -\theta ^{\prime })
\end{eqnarray}
In other words the two operators $\hat{Z}^{\#}(\theta )$ and $Z^{\#}(\theta
^{\prime })$ have relative canonical commutation relations which in turn
leads to the relative commutativity 
\begin{equation}
\left[ \hat{F}(\hat{f}),F(g)\right] =0,\,\,supp\hat{f}\in W^{opp},\,suppg\in
W
\end{equation}
The $F$ and $\hat{F}$ PFG's generate algebras $\mathcal{A}(W)$ and $\mathcal{%
A}(W)^{\prime }=alg\left\{ \hat{F}(\hat{f});supp\hat{f}\in W^{opp}\right\} =J%
\mathcal{A}(W)J$ and one easily checks 
\begin{eqnarray}
J\Delta ^{\frac{1}{2}}F(f_{1})...F(f_{n})\Omega
&=&(F(f_{1})...F(f_{n}))^{\ast }\Omega \\
\Delta ^{it} &=&U(\Lambda (2\pi i))  \notag
\end{eqnarray}
which is the defining relation for the Tomita operator $S=J\Delta ^{\frac{1}{%
2}}.$

The KMS computation can be extended to ``formfactors'' i.e. mixed
correlation functions containing in addition to F's one generic operator $%
A\in \mathcal{A}(W)$ so that the previous calculation results from the
specialization $A=1.$ This is so because the connected parts of the mixed
correlation function is related to the various $\left( n,m\right) $
formfactors obtained by the different ways of distributing n+m particles in
and out states using the relation between $Z^{\prime }s$ and Fock space
creation and annihilation operators. These different formfactors are
described by different boundary values of one analytic master function which
is in turn related to the various forward/backward on shell values which
appear in one mixed A-F correlation function. We may start from the
correlation function with one $A$ to the left and say n F's to the right and
write the KMS condition as 
\begin{equation}
\left\langle AF(\hat{f}_{n})...F(\hat{f}_{2})F(\hat{f}_{1})\right\rangle
=\left\langle F(\hat{f}_{1}^{2\pi i})AF(\hat{f}_{n})...F(\hat{f}%
_{2})\right\rangle  \label{A-KMS}
\end{equation}
The n-fold application of the F's to the vacuum on the left hand side
creates besides an n-particle term involving n operators $Z^{\ast }$ to the
vacuum (or KMS reference state vector) $\Omega $ also contributions from a
lower number of $Z^{\ast \prime }s$ together with $Z$-$Z^{\ast }$
contractions. As with free fields, the n-particle contribution can be
isolated by Wick-ordering\footnote{%
Note that as a result of the Z-F commutation relation the change of order
within the Wick-ordered products will produce rapidity dependent factors} 
\begin{equation}
\left\langle A:F(\hat{f}_{n})...F(\hat{f}_{2})F(\hat{f}_{1}):\right\rangle
=\left\langle F(\hat{f}_{1}^{2\pi i})A:F(\hat{f}_{n})...F(\hat{f}%
_{2}):\right\rangle  \label{KMS}
\end{equation}
Rewritten in terms of \ $Z^{\prime }s$ and using the denseness of the $%
f^{\prime }s$ this relation reads 
\begin{eqnarray}
&&\left\langle \Omega ,AZ^{\ast }(\theta _{n})...Z^{\ast }(\theta
_{2})Z^{\ast }(\theta _{1}-2\pi i)\Omega \right\rangle \\
&=&\left\langle \Omega ,Z(\theta _{1}+i\pi )AZ^{\ast }(\theta
_{n})...Z^{\ast }(\theta _{2})\Omega \right\rangle  \notag \\
&=&\left\langle Z^{\ast }(\theta _{1}-i\pi )\Omega ,AZ^{\ast }(\theta
_{n})...Z^{\ast }(\theta _{2})Z^{\ast }(\theta )\Omega \right\rangle  \notag
\end{eqnarray}
The analytic continuation by $2\pi i$ refers to the correlation function and
not to the operators. For the natural order of rapidities $\theta
_{n}>..>\theta _{1}$ this yields the following crossing relation (assuming
absence of boundstates) 
\begin{eqnarray}
&&\left\langle \Omega ,Aa_{in}^{\ast }(\theta _{n})...a_{in}^{\ast }(\theta
_{2})a_{in}^{\ast }(\theta _{1}-\pi i)\Omega \right\rangle \\
&=&\left\langle a_{out}^{\ast }(\theta _{1})\Omega ,Aa_{in}^{\ast }(\theta
_{n})...a_{in}^{\ast }(\theta _{2})\Omega \right\rangle  \notag
\end{eqnarray}
The out scattering notation on the bra-vectors becomes only relevant upon
iteration of the KMS condition since the bra $Z^{\prime }s$ have the
opposite natural order. By iteration one finally obtains the general mixed
matrix elements 
\begin{equation}
\left\langle a_{out}^{\ast }(\theta _{k})...a_{out}^{\ast }(\theta
_{1})\Omega ,Aa_{in}^{\ast }(\theta _{n})...a_{in}^{\ast }(\theta
_{k-1})\Omega \right\rangle  \label{formfactors}
\end{equation}
as analytic continuations from $\left\langle \Omega ,AZ^{\ast }(\theta
_{n})...Z^{\ast }(\theta _{2})Z^{\ast }(\theta _{1})\Omega \right\rangle $
which a posteriori justifies the use of the name formfactors in connection
with the mixed A-F correlation functions.

The upshot of this is that such an $A$ must be of the form 
\begin{equation}
A=\sum \frac{1}{n!}\int_{C}...\int_{C}a_{n}(\theta _{1},...\theta
_{n}):Z(\theta _{1})...Z(\theta _{n}):  \label{series}
\end{equation}
where the $a_{n}$ have a simple relation to the various formfactors of $A$
(including bound states) whose different in-out distributions of momenta
correspond to the different contributions to the integral from the
upper/lower rim of the strip bounded by C consisting of two contributions,
which are related by crossing. The transcription of the $a_{n}$ coefficient
functions into physical formfactors (\ref{formfactors}) complicates the
notation, since in the presence of bound states there is a larger number of
Fock space particle creation operators than the initial PFG wedge generators 
$F.$ It is comforting to know that the wedge generators, despite their lack
of vacuum polarization clouds, nevertheless contain the full (bound state)
particle content. The wedge algebra structure for factorizing models is like
a relativistic QM, but as soon as one sharpens the localization beyond wedge
localization, the field theoretic vacuum structure will destroy this simple
picture and replace it with the appearance of the characteristic virtual
particle structure which separates local quantum physics from quantum
mechanics.

In order to see by what mechanism the quantum mechanical picture is lost in
the next step of localization, let us consider the construction of the
double cone algebras as a relative commutants of shifted wedge (shifts by $a$
inside the standard wedge) 
\begin{eqnarray}
\mathcal{A}(C_{a}) &:&=\mathcal{A}(W_{a})^{\prime }\cap \mathcal{A}(W)
\label{loc} \\
C_{a} &=&W_{a}^{opp}\cap W  \notag
\end{eqnarray}
For $A\in \mathcal{A}(C_{a})\subset \mathcal{A}(W)$ and $F_{a}(\hat{f}%
_{i})\in \mathcal{A}(W_{a})\subset \mathcal{A}(W)$ the KMS condition for the
W-localization reads as before, except that whenever a $F_{a}(\hat{f}_{i})$
is crossed to the left side of $A,$ we may commute it back to the right side
since $\left[ \mathcal{A}(C_{a}),F_{a}(\hat{f}_{i})\right] =0.$ The new
relation resulting from the compact localization of $A$ is 
\begin{eqnarray}
&&\left\langle AF_{a}(\hat{f}_{1}):F_{a}(\hat{f}_{n})...F_{a}(\hat{f}%
_{2}):\right\rangle  \label{rec} \\
&=&\left\langle A:F_{a}(\hat{f}_{n})...F_{a}(\hat{f}_{2})F_{a}(\hat{f}%
_{1}^{2\pi i}):\right\rangle  \notag
\end{eqnarray}
Note that the $F_{a}(\hat{f}_{1})$ in the first line is outside the
Wick-ordering. Since it does neither act on the bra nor the ket vacuum, it
contains both frequency parts. The creation part can be combined with the
other $F$'s under one common Wick-ordering whereas the annihilation part via
contraction with one of the Wick-ordered $F$'s will give an expectation
value of one $A$ with $(n-2$) $F$'s. Using the density of the $f$'s and
going to rapidity space we obtain (\cite{S-W1}) the so-called kinematical
pole relation \cite{BFKZ} 
\begin{equation}
Res_{\theta _{12}=i\pi }\left\langle AZ^{\ast }(\theta _{n})...Z^{\ast
}(\theta _{2})Z^{\ast }(\theta _{1})\right\rangle =2i\mathbf{C}%
_{12}\left\langle AZ^{\ast }(\theta _{n})...Z^{\ast }(\theta
_{3})\right\rangle (1-S_{1n}...S_{13})  \label{pole}
\end{equation}
Here the product of two-particle S-matrices results from commuting the $%
Z(\theta _{1})$ to the right so that it stands to the left of $Z^{\ast
}(\theta _{2}),$ whereas the charge conjugation matrix $\mathbf{C}$ only
appears if we relax our assumption of selfcongugacy.

It is remarkable that this kinematical pole relation does not contain the
size of the localization region for $A.$ It is a relation which
characterizes all operator spaces $\mathcal{A}(\mathcal{O}),$ $\mathcal{O}%
\in W$ down to the pointlike limits. The individual localization sizes only
influence the Payley-Wiener exponents in asymptotic imaginary repidity
directions.

The existence problem for the QFT associated with an admissable S-matrix
(unitary, crossing symmetric, correct physical residua at one-particle
poles) of a factorizing theory is the nontriviality of the relative
commutant algebra i.e. $\mathcal{A}(C_{a})\neq \mathbf{C}\cdot 1.$
Intuitively the operators in double cone algebras are expected to behave
similar to pointlike fields applied to the vacuum; namely one expects the
full interacting polarization cloud structure. For the case at hand this is
in fact a consequence of the above kinematical pole formula since this
formula leads to a recursion which for nontrivial two-particle S-matrices is
inconsistent with a finite number of terms in (\ref{series}). Only if the
bracket containing the S-products vanishes, the operator $A$ is a composite
of a free field.

The modular method has therefore converted the existence problem, which
hitherto was dominated by the well-known ultraviolet behavior of special
(Lagrangian) field-coordinates, into the problem of nontriviality of
algebraic intersections or in more applied terms to the nontriviality of
formfactor spaces. For special fields which have an intrinsic meaning as
conserved currents and their related order/disorder structure (example: the
conserved current and its Sine-Gordon potential in the massive Thirring
model) one expects to be able to identify them individually and to compute
their formfactors as well as their correlation function. The considerations
in the next section will propose arguments that this modular construction
method is not limited to factorizing models.

The determination of a relative commutant or an intersection of wedge
algebras is even in the context of factorizing models not an easy matter. We
expect that the use of the following ``holographic'' structure significantly
simplifies this problem. We first perform a \textit{lightlike translation}
of the wedge into itself by letting it slide along the upper light ray by
the amount given by the lightlike vector $a_{+}.$ We obtain an inclusion of
algebras and an associated relative commutant 
\begin{eqnarray}
&&\mathcal{A}(W_{a_{\pm }})\subset \mathcal{A}(W) \\
&&\mathcal{A}(W_{a_{\pm }})^{\prime }\cap \mathcal{A}(W)  \notag
\end{eqnarray}
The intuitive picture is that the relative commutant lives on the $a_{\pm 
\text{ }}$interval of the upper/lower light ray, since this is the only
region inside W which is spacelike to the interior of the respective shifted
wedges. This relative commutant subalgebra is a light ray part of the above
double cone algebra, and it has an easier mathematical structure. One only
has to take a generic operator in the wedge algebra which formally can be
written as a power series in the generators $Z$ and find those operators 
\cite{Sch1}\cite{S-W2} which commute with the shifted F's 
\begin{equation}
\left[ A,U(e_{+})F(f)U^{\ast }(e_{+})\right] =0
\end{equation}
Since the shifted F's are linear expressions in the Z's, the $n^{th}$ order
polynomial contribution to the commutator comes from only two adjacent terms
in $A$ namely from $a_{n+1}$ and $a_{n-1}$ which correspond to the
annihilation/creation term in F. The result is precisely the same as the one
from the KMS property: the above kinematical pole formula (\ref{pole}), so
we do not learn anything new beyond what was already observed with the KMS
technique. However as will be explained below, the net obtained from the
algebra 
\begin{equation}
\mathcal{A}(\mathbf{R}_{\pm }):=\vee _{b_{\pm }}AdU(b_{\pm })\left\{ \vee
_{a_{\pm }}\mathcal{A}(W_{a\pm })^{\prime }\cap \mathcal{A}(W)\right\}
\end{equation}
(in words the net of von Neumann algebras created by translating the
relative commutants of size $a_{\pm }$ with $b_{\pm }$ along the upper/lower
light rays) is a chiral conformal net on the respective subspace $H_{\pm }=$ 
$\overline{\mathcal{A}_{\pm }\Omega }$ which is indexed by intervals on the
light ray$.$ If our initial algebra were d=1+1 conformal theories, the total
space would factorize $H=H_{+}\bar{\otimes}H_{-}=$ $\overline{\left( 
\mathcal{A}_{+}\bar{\otimes}\mathcal{A}_{-}\right) \Omega },$ and we would
recover the well-known fact that two-dimensional local theories factorize
into the two light ray theories. For massive theories we expect $H=\overline{%
\mathcal{A}_{+}\Omega }=\overline{\mathcal{A}_{-}\Omega },$ i.e. the Hilbert
space obtained from one light ray horizon already contains all state vectors$%
.$ This would correspond to the difference in classical propagation of
characteristic massless versus massive data in d=1+1. There it is known that
although for the massless case one needs the characteristic data on the two
light rays, the massive case requires only one light ray. In fact there
exists a rigorous proof that this classical behavior carries over to free
quantum fields: with the exception of m=0 massless theories, in all other
cases (including light-front data for higher dimensional m=0 situations) the
vacuum is cyclic with respect to one light front $H=\overline{\mathcal{A}%
_{\pm }\Omega }$ \cite{GLRV}. The proof is representation-theoretical and
holds for all cases except the d=1+1 massless case. The result may be
written as an identity of global algebras 
\begin{equation}
\mathcal{A}(W)=\mathcal{A}(\mathbf{R}_{+}^{>})
\end{equation}
where the superscript refers to the fact that we are considering the right
half of the upper light ray (with the same relation for the lower light
ray). This identity of global algebras, which we consider as an AQFT version
of holography, does not extend to the natural net structure which consists
of double cones in W resp. intervals on \textbf{R}$_{+}^{>}.$ This means
that certain geometric actions as the lower light cone translation $U(a_{-})$
on the W-net will be extremely nonlocal in their action on $\mathcal{A}(%
\mathbf{R}_{+}^{>}).$ The appearance of these ``hidden symmetries'' is the
prize one has to pay for the simplifications of lower dimensional
holographic images. More remarks on holography for higher dimensional QFT
can be found in a later part.

It almost goes without saying that the various restrictions we have imposed
for pedagogical reasons on the Z-algebra structure (as diagonal structure of
S and absence of poles) can easily be lifted.

\subsection{Case with Real Particle Creation}

For models with real particle creation it is not immediately clear how to
construct PFG's, in fact it is not obvious whether they exist. On the other
hand it is quite easy to see that for any smaller localization region (whose
causal completion will not be as big as a wedge) there can be no PFG-like
operators unless the theory is trivial (i.e. free in the sense of no
interaction). This means that PFG's are ideal indicators for interactions
because only polarization caused by interactions will appear\footnote{%
The vacuum polarization clouds which are responsible for the localization
entropy in the later section are also present in the free case.}. With other
words any operator with compact or even spacelike cone localization which
couples the one-particle state with the vacuum if applied once to the vacuum
will generate a polarization cloud on top of the one-particle state unless
the particles are noninteracting. The proof of this theorem uses similar
analytic techniques as that of the Jost-Schroer theorem\cite{Wightman}. In
fact the proof follows almost literally the arguments of Mund \cite{Mund}
where these analytic techniques were recently used to show that the d=2+1
braid-group particles even in their ``freest'' form cannot be quantum
mechanical objects i.e. they cannot be described by localized operators
which carry a defined (incoming) particle number like free Bosons/Fermions
and hence a nonrelativistic limit which maintains the plektonic
spin-statistic connection will also maintain the vacuum polarization
structure and hence be outside of quantum mechanics. In terms of a
representation-theoretical setting of multi-particle states one looses the
tensor product structure of n-particle scattering states in terms if Wigner
one-particle states. For a more remarks on the ``No-Go theorem for
interacting PFG's with smaller than wedge localization'' I refer to a
forthcoming paper \cite{BBS}.

An existence proof of wedge-localized PFG which as unbounded operators
associated with $\mathcal{A}(W)$ (i.e. the proper PFG's for the purpose of
this essay) is simple if one allows also unbounded PFG operators associated
with the von Neumann algebras can be given. One first studies the
wedge-localization spaces i.e. the vectors spanning the domain of $\Delta ^{%
\frac{1}{2}}$ which are the vectors in the thermal subspace $%
H_{R}(W)+iH_{R}(W)$ where $H_{R}(W)$ is the closed real subspace of
solutions of the localization equation 
\begin{equation}
S\psi =\psi ,\,\,S=J\Delta ^{\frac{1}{2}},\,J=S_{scat}J_{0}  \label{eigen}
\end{equation}
This space has a nontrivial intersection with the one-particle subspace 
\begin{equation}
H_{R}(W)\cap H_{Wigner}^{(1)}\neq 0
\end{equation}
which is a consequence of the fact that the modular operator $\Delta ^{it}$
is shared with that of the wedge algebra generated by the free asymptotic
(incoming) fields. The possibility of representing each vector as an
unbounded operator associated with $\mathcal{A}(W)$ is guarantied by modular
theory and this applies in particular to a dense set of one-particle vectors.

In order to get a clue for the construction of the spaces we look at d=1+1
theories which do not have any transversal extension to wedges. Furthermore
we assume that there is only one kind of particle (absence of particle poles
in the S-matrix) so that in terms of incoming particles one is in the
situation of a Fock space with one kind of particle.

From the previous discussion we take the idea that we should look for a
relation between the ordering of rapidities and the action of the scattering
operator. Therefore we define a subspace indexed by two-particle wave
functions as follows (omitting again the scat subscript): 
\begin{eqnarray}
\Psi _{f_{2},f_{1}} &\equiv &\int \int d\theta _{1}d\theta _{2}f_{2}(\theta
_{2})f_{1}(\theta _{1})\Psi (\theta _{2},\theta _{1}) \\
\Psi (\theta _{2},\theta _{1}) &\sim &\chi _{21}a^{\ast }(\theta
_{2})a^{\ast }(\theta _{1})\Omega +\chi _{21}Sa^{\ast }(\theta _{1})a^{\ast
}(\theta _{2})\Omega  \notag
\end{eqnarray}
It is easy to check that this vector fulfils (\ref{eigen}) if the f's have
the properties of the previous section. The $J_{0}$ sends the $S$ into a $%
S^{\ast }$ and the f's into their complex conjugate whereas the $S$ together
with the unitarity reproduces the linear combination. Finally the $\Delta ^{%
\frac{1}{2}}$ makes a $i\pi $ shift in the $\theta ^{\prime }s$ which may be
absorbed into the $f^{\ast \prime }s$ with the result that the original $%
f^{\prime }s$ are reproduced.

The generalization to states indexed by 3 $f^{\prime }s$ contain 6
contributions which correspond to the 6 permutations

\begin{multline}
\,\,\,\,\,\,\,\,\,\,\,\,\,\,\,\,\,\,\,\,\,\,\,\,\,\,\,\,\,\,\,\Psi
_{f_{3}f_{2}f_{1}}\equiv \int \int \int d\theta _{1}d\theta _{2}d\theta
_{3}f_{3}(\theta _{3})f_{2}(\theta _{2})f_{1}(\theta _{1})\Psi (\theta
_{3},\theta _{2},\theta _{1})  \label{3-p} \\
\Psi (\theta _{3},\theta _{2},\theta _{1})\symbol{126}\chi _{321}a^{\ast
}(\theta _{3})a^{\ast }(\theta _{2})a^{\ast }(\theta _{1})\Omega +\chi
_{312}S_{21}a^{\ast }(\theta _{3})a^{\ast }(\theta _{2})a^{\ast }(\theta
_{1})\Omega  \notag \\
+\chi _{231}S_{32}a^{\ast }(\theta _{3})a^{\ast }(\theta _{2})a^{\ast
}(\theta _{1})\Omega +\chi _{123}S_{321}a^{\ast }(\theta _{3})a^{\ast
}(\theta _{2})a^{\ast }(\theta _{1})\Omega  \notag \\
+\chi _{132}S_{321}\cdot S_{23}^{\ast }a^{\ast }(\theta _{3})a^{\ast
}(\theta _{2})a^{\ast }(\theta _{1})\Omega +\chi _{213}S_{321}\cdot
S_{12}^{\ast }a^{\ast }(\theta _{3})a^{\ast }(\theta _{2})a^{\ast }(\theta
_{1})\Omega  \notag
\end{multline}
This expression results from writing each permutation as the nonoverlapping
product of ``mirror permutations''. The smallest mirror permutations are
transpositions of adjacent factors as in the third and fourth term. For
those one replaces the action of the permutation by the action of the
S-matrix restricted to the adjacent transposed tensor factors (which is used
as a subscript of S). An example for an overlapping product of
transpositions is the product of two transpositions which have one element
in common e.g. $123\rightarrow 132\rightarrow 312;$ this sequence of mirror
permutations can not be associated with subsequent S-matrix actions on
tensor products. However the composition $123\rightarrow 213\rightarrow 312$
has a meaningful S-matrix counterpart: namely $S\cdot S_{12}a^{\ast }(\theta
_{1})a^{\ast }(\theta _{2})a^{\ast }(\theta _{3})\Omega $ where $S_{12}$
leaves the third tensor factor unchanged. The resulting vector under the $%
S_{12}$ action has no well-defined incoming particle number and can also be
written in tensor product notation as $\left( Sa^{\ast }(\theta _{1})a^{\ast
}(\theta _{2})\Omega \right) \otimes a^{\ast }(\theta _{3})\Omega .$ The
third particle has remained a spectator and only enters the process when the
final $S$ is applied (which corresponds to the mirror permutation of all 3
objects). This action of nested mirror transformations is well-defined. In
general if one mirror permutation is completely inside a larger one the
scattering corresponding nested product of $S^{\prime }s$ is a well-defined
physical meaning. The last two terms correspond are such nested mirror
contributions. The inner products of such vectors with themselves will lead
to matrix elements of the form

\begin{equation}
\left\langle \theta _{3}^{\prime },\theta _{2}^{\prime },\theta _{1}^{\prime
}\left| S\cdot S_{12}^{\ast }\right| \theta _{3},\theta _{2},\theta
_{1}\right\rangle   \label{dot}
\end{equation}
In a graphical scattering representation particle 1 and 2 would scatter
first and produce arbitrarily many particles (subject to the conservation
laws for the total energy-momentum) which together with the third incoming
particle (which hitherto was only a spectator) enter an additional
scattering process of which only the 3-particle outgoing component is
separated out by the matrix element in (\ref{dot}). The dot means summation
over all admissable intermediate states and could be represented by e.g. a
heavy line in the graphical representation in order to distinguish it from
the one-particle lines. Whereas in the calculation of cross sections the
summation over intermediate states lead to diagonal inclusive processes, the
nested structure of the localized vectors correspond to non-diagonal
inclusive processes. The proof that the space of vectors of the above form $%
\Psi _{f_{n}...f_{1}}$ fulfil (\ref{eigen}) is analogous to the previous
case: the first and the fourth term change their role as well as the second
and third terms change role with the two nested terms.

For a 4- $f$ labeled state vector $\Psi _{f_{4}f_{3}f_{2}f_{1}}$ there is
the new possibility of having two -particle S's acting on two nonoverlapping
pairs of in-particles before the action of either the identity or the full
S-matrix is applied. For further details we refer to \cite{S2}. The full
real wedge localization space is defined as the real closure of all the
labeled spaces (labeled by wedge localized one-particle wave functions) 
\begin{equation}
H_{R}(W)=real\,closure\left\{ \Psi _{f},\Psi _{f_{2}f_{1}},\Psi
_{f_{3}f_{2}f_{1}},\Psi _{f_{4}f_{3}f_{2}f_{1}},...|\forall f_{i}\in
H^{(1)}(W)\right\}
\end{equation}

The remaining problem is whether one can generate the wedge localization
spaces by the iterated application of PFG operators. The check of the
equivalence between KMS and on-shell crossing symmetry would then proceed as
before by forming inner products between these vectors. The understanding of
the precise mathematical status of these PFG's was still an open problem at
the time of writing. It is clear that in the case of real particle creation
one looses the uniformization aspect in the rapidity in which the S-matrix
and formfactors were meromorphic functions. The distribution theoretical
aspect of functions with infinitely many piled-up cuts on the real rapidity
axis i.e. their localization in rapidity space may cause problems (in the
above expressions they are integrated with boundary values of analytic wave
functions $f$). For more informations we again refer to a forthcoming
publication \cite{BBS}.

There are several reasons why constructions based on modular localization
could be important for particle physics. Besides the improvement in the
understanding the structure of interacting QFT one expects that they could
lead to a perturbation theory of local nets which bypasses the use of the
nonintrinsic field coordinatizations and also the appearance of
short-distance ultraviolet divergencies. The perturbative construction of
vacuum expectations of PFG's which generate wedge algebras is reminiscent to
a the revival of the perturbative version of the old dream to construct an
S-matrix just using crossing symmetry (and the analyticity which is required
for its formulation) in addition to unitarity. The old S-matrix bootstrap
program failed, even on a perturbative level no formulation without the use
of fields was found. But thanks to modular wedge localization we can now
formulate a similar but structurally richer program which already showed its
power in the case of factorizing models. It is clear now that the weak point
of the old S-matrix bootstrap was not primarily in its concepts but rather
in its almost ideological and unfounded stance against QFT and anything
``off-shell''. For a recent review of S-matrix theory I refer to \cite{White}%
. Finally the claim that it is a unique theory and that it constituted a
``TOE'' (a theory of everything, in this case everything minus quantum
gravity) contributed to its downfall. The present modular localization
approach is different on all counts. Even the avoidance of field
coordinatizations in favor of nets has entirely pragmatic reasons. In
sharpening the localization beyond wedges via algebraic intersections of
wedge algebras instead of using the local coupling of fields with its short
distance problems and rather ad hoc resulting separation into
renormalizable/nonrenormalizable, one has the chance to shed an entirely new
light on problems which are central to QFT.

\subsection{Modular Origin of Quantum Symmetries}

Modular theory reproduces all the standard spacetime and internal
symmetries, but it also produces new symmetries which remained hidden to the
Lagrangian approach.

Before we look at the hidden symmetries, it is interesting to note that even
the standard symmetries (i.e. those having a classical Noetherian
counterpart) reappear in a very unusual and interesting way. To illustrate
this point let us ask how can we characterize a chiral conformal theory i.e.
its algebraic description in terms of a net on the circle. The well-known
answer is: \textit{by two algebras which are in the relative position of ``
half-sided modular inclusion'' (hsm)} \cite{Wies}. The prototype are two
half-circle algebras rotated by $\frac{\pi }{2}$ relative to each other (the
quarter-circle situation) \cite{S}. The $\frac{1}{4}$-circle of their
intersection is compressed towards one of its endpoints under the action of
each of the dilations associated with the half-circle which are the modular
groups of the associated algebras. In fact the compression only happens for
one particular ($\pm $)sign of the dilation parameter ($\pm hsm$). This
together with the analytic results by Borchers coming from the energy
positivity within the modular setting \cite{Bo}, inspired Wiesbrock to
introduce a general theory of modular inclusions and modular intersection.
With respect to chiral conformal theories Wiesbrock's result was that the
study of abstract ``standard hsm-inclusions'' is equivalent to the
classification of chiral conformal nets.

Encouraged by this success, this modular inclusion concept was enriched by
additional requirement of a more geometric nature whereupon it became
possible to characterize also higher dimensional nonconformal nets in terms
of the modular relations (inclusions, intersections) of a finite family of
von Neumann algebras. The surprising aspects of these investigation was that
both the spacetime symmetries (the Poincar\'{e} or conformal symmetries) as
well as the physics-encoding net structure follow from abstract relations
(modular inclusions, intersections) between a finite number of copies of one
and the same unique von Neumann algebra (the hyperfinite III$_{1}$-factor).
In view of the fact that the modular groups of most causally complete
regions act as unknown non-pointlike transformations, it was interesting to
get more information about their interpretation in terms of physical
symmetries \cite{S-W1}. Again it appeared reasonable to study of this
question in the simplest context of chiral conformal theories. In contrast
to higher dimensions chiral theories do have infinitely many geometrically
acting one-parametric diffeomorphisms which are unitarily implemented by
unitaries which change the vacuum. It turns out that by taking the large
parameter limit (see next section for an example) the transformed
correlation functions stabilize and define a new state over the algebra
which is invariant under the respective subgroup. A closer examination
reveals that these states have a modular interpretation with respect to
multi-interval algebras which are cyclic and separating with respect to this
state (but loose this cyclicity upon restriction to one algebra). This
explains the modular aspects of all spacetime regions on the circle,
including disconnected ones. By contrast, in higher dimensions the modular
groups of massive theories (with the exception of wedge regions) are for no
choice of states pointlike\footnote{%
The best one can hope for is that they act asymptoticall pointlike near the
causal horizon.}; they preserve the causal closure of the localization
region but act nonlocally inside (they would act on localizing test
functions in a support-preserving but otherwise nonlocal fuzzy way). By
analogy one should then view a suitably defined universal infinite
parametric modular group generated by all the individual modular groups of
spacetime regions as the hidden symmetry analogue of the chiral
diffeomorphism group. The Poincar\'{e} group is the maximal geometric
subgroup and it is generated from a finite subset of ($\mathcal{A}(W),\Omega
)$ $W\in \mathcal{W}$. One also meets ``partially hidden'' symmetries in the
spacetime analysis of modular inclusions/intersections i.e. automorphisms
which act geometrically on subnets.

The present method of analysis based on modular groups is not the only one;
a very interesting alternative approach based on the modular involutions $J$
has been proposed by \cite{BDFS}.

Closely related to the issue of hidden symmetries is the inverse of the
Unruh observation namely the question of existence of a geometrical
interpretation ``behind the horizon'' of the von Neumann commutant of a
thermal heat bath system. Conditions under which this is possible have been
studied in \cite{S-W3}\cite{B-Y}

The reduction of LQP to the study of inclusions and intersections has
changed the underlying philosophical basis of particle physics. The
different outlook had been occasionally described by Haag in terms of a
change from the Newtonian picture of reality as a manifold filled with a
material content (relativity and quantum mechanics included) to the world of
monades of Leibnitz, which although lacking individuality, create a rich
reality by their interrelations.

The reader is invited to try to translate Leibnitz'es monades into
hyperfinite type III$_{1}$ von Neumann factors. The latter are as
structureless entities and like points in geometry without individuality
with one important difference: one factor can be included in the other and
both can have nontrivial intersection. One should mention that this mode of
thinking is also quite visible in the mathematics discovered by Alain Connes
and in Vaugn Jones subfactor theory.

\section{Local Quantum Physics versus Quantum Mechanics: a Change of Paradigm%
}

The consequences of modular localization as explained in the previous
section are not the only source of radical conceptual change in QFT. Another
equally drastic conceptual change change of paradigm (however with a strict
adherence to the physical principles of LQP) is the ``degree of freedom'' or
phase space property of QFT and the positioning of QM versus QFT.

\subsection{The LQP Phase Space}

Again this has a quite interesting history behind it, although some of its
more dramatic consequences were only noticed in more recent times. It goes
back to attempts by Haag and Swieca to make some of the consequences of the
density of local states as expressed in the Reeh-Schlieder density theorem%
\footnote{%
The Reeh-Schlieder denseness theorem \cite{Wightman} is often presented
together with the assertion of a one-to one correspondence between localized
operators and vectors in the dense subspace of localized states, the so
called separability property (the ``operator-state correspondence'') .
Modular theory allows a profound understanding and relates denseness and
separability as dual properties in the sense of von Neumann's commutant
notion.} more physically acceptable by introducing additional concepts \cite
{Haag}. Whereas in quantum mechanics the box localization separates the
physical description via tensor-product factorization into an ``inside and
outside Hilbert space'' (and a corresponding tensor-product of full operator
algebras), the long range vacuum structure due to the omnipresence of vacuum
fluctuations destroys such a picture and replaces it by an extreme opposite
denseness (cyclicity) property of localized state vectors, the so-called
Reeh-Schlieder property. This denseness property of localized states 
\begin{equation*}
\overline{\mathcal{A}(\mathcal{O})\Omega }=H
\end{equation*}
has been sometimes provocatively referred to by some of the protagonists of
these investigations as the ``particle creation behind the moon''-paradox:
by applying appropiate observables localized in spacetime to the vacuum one
may approximate any local change anywhere instantaneously. Even if one (as
one learned from the analysis of ERP Gedankenexperiment) is prepared to make
a distinction between causal ties of events and long range correlations in
states, this does not explain why there is such an impressive conceptual
difference between the tensor factorization of quantum mechanical
localization and the localization in LQP.

In an attempt to reconcile the strange-looking aspects with common sense in
quantum theory, Haag and Swieca introduced the notion of phase space into
LQP. They restricted the local vector states by the requirement that $P_{E}%
\mathcal{A}^{(1)}(\mathcal{O})\Omega \,\,$be\ a compact set of vectors in $H$%
. Here the superscript on $\mathcal{A}(\mathcal{O})$ denotes the unit ball
in the operator norm of the local algebra and $P_{E}$ is the projector on
vectors of energy smaller than E which feature in the spectral
representation of the hamiltonian $\mathbf{H}=\int EdP_{E}.$ They argue that
the creation of ``behind the moon states'' in an earthly laboratory is not
possible with a limited supply of energy i.e. the incredible small vacuum
polarization correlations which exist as a matter of principle even over
large distances can not be sufficiently amplified in the desired region with
a limited energy supply. Using the same type of intuition but sharper
estimates, Buchholz and Wichmann proposed a variant of this requirement
which became known under the name nuclearity requirement and has the
advantage that it is easier to use in calculations and closer to properties
of thermal states. It reads 
\begin{equation}
P_{E}\mathcal{A}(\mathcal{O})\Omega \,\,or\,\,e^{-\beta \mathbf{H}}\mathcal{A%
}(\mathcal{O})\Omega \;is\,\,nuclear
\end{equation}
This amounts to the nuclearity of the map $\Theta :$ $\mathcal{A}(\mathcal{O}%
)\rightarrow e^{-\beta H}\mathcal{A}(\mathcal{O})\Omega $ \ i.e. the
requirement that this map has a representation 
\begin{equation}
\Theta A=\sum \phi _{i}(A)\psi _{i}
\end{equation}
where the $\phi _{i}$ are bounded linear forms on the algebra and the $\psi
_{i}$ are vectors in the Hilbert space such that 
\begin{eqnarray}
&&\sum \left\| \phi _{i}\right\| \left\| \psi _{i}\right\| <\infty \\
&&\left\| \Theta \right\| _{1}:=inf\sum \phi _{i}(A)\psi _{i}
\end{eqnarray}
with the norms having the respective natural meaning and the last equation
defines a new ``nuclear norm'' \cite{Haag}. The requirement implies that the
image set in the Hilbert space is ``nuclear'' and a fortiori compact as
demanded by Haag-Swieca. In physics terms such maps are only nuclear if the
mass spectrum of LQP is not too accumulative in finite mass intervals; the
excluded cases are those which in quantum statistical mechanics would cause
the strange appearance of a maximal ``Hagedorn'' temperature or the complete
loss of thermal concepts, so that one expects a close relation between
nuclearity and the thermal aspects of QFT. Indeed the nuclearity assures
that a QFT, which was given in terms of its vacuum representation, also
exists in a thermal state. In fact the nuclearity index turns out to be the
counterpart of the quantum mechanical Gibbs partition function \cite{Bu}\cite
{Haag} for open systems and behaves in an entirely analogous way to the
Gibbs formula in a closed quantization box. The nuclearity property and the
resulting phase space properties of LQP (localization in spacetime and
limitation of energy) goes a long way to reconcile the local denseness of
state property with common sense in that it associates with an approximating
sequence of ``particle behind the moon creation'' an ever increasing
expenditure in energy.

\subsection{The Split Property}

Before we link nuclearity with the pivotal ``split property'', let us
motivate the latter taking a helping hand from the history of QFT. The
peculiarities of the above degrees-of-freedom-counting are very much related
to one of the oldest ``exotic'' and at the same time characteristic aspects
of QFT, namely vacuum polarization. As first noticed by Heisenberg (and
later elaborated and used by Euler, Weisskopf and many others), the partial
charge: 
\begin{equation}
Q_{V}=\int_{V}j_{0}(x)d^{3}x=\infty
\end{equation}
diverges as a result of uncontrolled vacuum particle/antiparticle
fluctuations near the boundary. In order to quantify this divergence one
acts with more carefully defined partial charges on the vacuum (s=dimension
of space) 
\begin{equation}
Q_{R}=\int j_{0}(x)f(x_{0})g(\frac{\mathbf{x}}{R})d^{s}x
\end{equation}
The vectors $Q_{R}\Omega $ only converge weakly for R$\rightarrow \infty $
on a dense domain. Their norms diverge as \cite{BDLR} 
\begin{eqnarray}
\left( Q_{R}\Omega ,Q_{R}\Omega \right) &\leq &const\cdot R^{s-1}
\label{surface} \\
&\sim &area  \notag
\end{eqnarray}
The surface character of this vacuum polarization is reflected in the area
behavior. Different from the vacuum polarization clouds in the previous
sections this surface vacuum polarization exists even without interactions.

The algebraic counterpart of this age-old observation is the so called
``split property'', namely the statement \cite{Haag} that if one leaves
between say the double cone (the inside of a ``relativistic box'')
observable algebra $\mathcal{A}(\mathcal{O})\,$and its causal disjoint (its
relativistic outside) $\mathcal{A}(\mathcal{O}^{\prime })$ a ``collar''
(geometrical picture of the relative commutant) $\mathcal{O}_{1}^{\prime
}\cap \mathcal{O}$, i.e. 
\begin{equation}
\mathcal{A}(\mathcal{O})\subset \mathcal{A}(\mathcal{O}_{1}),\,\,\,\mathcal{%
O\ll O}_{1}\,,\,\,properly
\end{equation}
then it is possible to construct in a canonical way a type $I$ tensor factor 
$\mathcal{N}$ which extends in a ``fuzzy'' manner into the collar $\mathcal{A%
}(\mathcal{O})^{\prime }\cap \mathcal{A}(\mathcal{O}_{1})$ i.e. $\mathcal{A}(%
\mathcal{O})\subset \mathcal{N}\subset \mathcal{A}(\mathcal{O}_{1}).$ With
respect to $\mathcal{N}$ the Hilbert space factorizes i.e. as in QM there
are states with no fluctuations (or no entanglement) for the ``smoothened''
operators in $\mathcal{N}.$ Whereas the original vacuum will be entangled
from the box point of view, there also exists a disentangled product vacuum
on $\mathcal{N}.$ The algebraic analogue of a smoothening of the boundary by
a test function is the construction of a factorization of the vacuum with
respect to a suitably constructed type $I$ factor algebra which uses the
above collar extension of $\mathcal{A}(\mathcal{O}).$ It turns out that
there is a canonical, i.e. mathematically distinguished factorization, which
lends itself to define a natural ``localizing map'' $\Phi $ and which has
given valuable insight into an intrinsic LQP version of Noether's theorem 
\cite{Haag}, i.e. one which does not rely on quantizing classical Noether
currents. It is this ``split inclusion'' which allows to bring back the
familiar structure of pure states, tensor product factorization,
entanglement and all the other properties at the heart of standard quantum
theory and the measurement process. However despite all the efforts to
return to structures known from QM, the original vacuum retains its thermal
(entanglement) properties with respect to all localized algebras, even with
respect to the ``fuzzy'' localized $\mathcal{N}.$

Let us collect in the following some useful mathematical definitions and
formulas for ``standard split inclusions'' \cite{Do-Lo}

\textit{Def.}: An inclusion $\Lambda =(\mathcal{A},\mathcal{B},\Omega )$ of
factors is called standard split if the collar $\mathcal{A}\prime \cap 
\mathcal{B}$ as well as $\mathcal{A},\mathcal{B}$ together with $\Omega $
are standard in the previous sense, and if in addition it is possible to
place a type I$_{\infty }$ factor $\mathcal{N}$ between $\mathcal{A}$ and $%
\mathcal{B}$.

In this situation there exists a canonical isomorphism of $\mathcal{A}\vee 
\mathcal{B}^{\prime }$ to the tensor product $\mathcal{A}\bar{\otimes}%
\mathcal{B}^{\prime }$ which is implemented by a unitary $U(\Lambda
):H_{\Lambda }\rightarrow H_{1}\bar{\otimes}H_{2}$ (the ``localizing map'')
with 
\begin{eqnarray}
&&U(\Lambda )(AB^{\prime })U^{\ast }(\Lambda )=A\bar{\otimes}B^{\prime } \\
&&A\in \mathcal{A},\,\,B^{\prime }\in \mathcal{B}^{\prime }  \notag \\
&&U^{\ast }(\Lambda )(\Omega \bar{\otimes}\Omega )\equiv \eta _{\Lambda }\in
H_{\Lambda }  \notag \\
&&\left\langle \eta _{\Lambda }\left| AB^{\prime }\right| \eta _{\Lambda
}\right\rangle =\omega (A)\omega (B^{\prime })\neq \omega (AB^{\prime }) 
\notag
\end{eqnarray}
This map permits to define a canonical intermediate type I factor $\mathcal{N%
}_{\Lambda }$ (which may differ from the $\mathcal{N}$ in the definition) 
\begin{equation}
\mathcal{N}_{\Lambda }:=U^{\ast }(\Lambda )B(H_{1})\otimes \mathbf{1}%
U(\Lambda )\subset \mathcal{B}\subset B(H_{\Lambda })
\end{equation}
It is possible to give an explicit formula for this canonical intermediate
algebra in terms of the modular conjugation $J=U^{\ast }(\Lambda )J_{%
\mathcal{A}}\otimes J_{\mathcal{B}}U(\Lambda )\,$\ of the collar algebra ($%
\mathcal{A}^{\prime }\cap \mathcal{B},\Omega )$ \cite{Do-Lo} 
\begin{equation}
\mathcal{N}_{\Lambda }=\mathcal{A}\vee J\mathcal{A}J=\mathcal{B}\wedge J%
\mathcal{B}J
\end{equation}
\thinspace

The tensor product representation gives the following equivalent tensor
product representation formulae for the various algebras 
\begin{eqnarray}
\mathcal{A} &\sim &\mathcal{A}\otimes \mathbf{1} \\
\mathcal{B}^{\prime } &\sim &\mathbf{1}\otimes \mathcal{B}^{\prime }  \notag
\\
N_{\Lambda } &\sim &B(H_{\Lambda })\otimes \mathbf{1}  \notag
\end{eqnarray}
As explained in \cite{Do-Lo}, the uniqueness of $U(\Lambda )$ and $%
N_{\Lambda }$ is achieved with the help of the ``natural cones'' $\mathcal{P}%
_{\Omega }(\mathcal{A}\vee \mathcal{B}^{\prime })$ and $\mathcal{P}_{\Omega
\otimes \Omega }(\mathcal{A}\otimes \mathcal{B}^{\prime }).$ These are cones
in Hilbert space whose position in $H_{\Lambda }$ together with their facial
subcone structures preempt the full algebra structure on a spatial level.
The corresponding marvelous theorem of Connes \cite{Connes} goes far beyond
the previously mentioned state vector/field relation which follows from the
Reeh-Schlieder density theorem.

Returning to our physical problem, we note that we have succeeded to find
the right analogue of the QM box for open LQP subsystems. Contrary to the
hyperfinite type $III_{1}$ algebras for causally closed double cone regions
with their sharp light cone boundaries (``quantum horizons''), the ``fuzzy
box'' type I factor $N_{\Lambda }$ constructed above (apart from its fuuzzy
geometrical aspects) permits all the properties we know from QM: pure
states, inside/outside tensor-factorization, (dis)entanglement etc. Whereas $%
\mathcal{A}$ as a type $III$ algebra is ``intrinsically entangled''\footnote{%
Such algebras have neither pure states nor can they appear as tensor-factors
in the factorization of bigger algebras. Their properties from the quantum
measurement point of view are nicely explained in \cite{C-H}.}, the fuzzy
box is a conventional quantum mechanical algebra whose only unusual aspect
is that the restriction of the vacuum generates entanglement and a
Hawking-Unruh temperature. Mathematically this means that the state $\omega
\mid _{\mathcal{A}\bar{\otimes}\mathcal{B}^{\prime }}$represented in the
tensor product cone $P_{\Omega \otimes \Omega }(\mathcal{A}\bar{\otimes}%
\mathcal{B}^{\prime })$ is not the tensor-product of those of the separate
restrictions of $\omega $ to $\mathcal{A}$ and $\mathcal{B}^{\prime }$ but
rather a highly entangled KMS temperature state. This is obviously the
result of vacuum fluctuations i.e. the fact that a physical vacuum in a LQP,
different from the no-particle state of Schr\"{o}dinger QM, correlates
spatially separated regions. Note also that the restriction of the product
state $\omega \otimes \omega $ to $\mathcal{B}$ or $\mathcal{B}^{\prime }$
is not faithful resp. cyclic on the corresponding vectors and therefore the
application of those algebras to the representative vectors $\eta _{\omega
\otimes \omega }$ yields projectors (e.g. $P_{\Lambda }=U^{\ast }(\Lambda
)B(H_{1})\bar{\otimes}1U(\Lambda )).$

\subsection{Localization-Entropy}

Since the fuzzy box algebra $N_{\Lambda }$ is of quantum mechanical type $I$%
, we are allowed to use the usual trace formalism based on the density
matrix description, i.e. the vacuum state is a highly entangled density
matrix $\rho _{\Omega }$ on $\mathcal{N}_{\Lambda }$ which leads to a
well-defined von Neumann entropy 
\begin{eqnarray}
\left( \Omega ,A\Omega \right) &=&tr\rho _{\Lambda }A,\,\,A\in \mathcal{A} \\
S(\rho _{\Lambda }) &=&-tr\rho _{\Lambda }log\rho _{\Lambda }
\end{eqnarray}
It turns out to be quite difficult to actually compute $\rho _{\Lambda }$
which describes the von Neumann entropy of the fuzzy box $S(\rho _{\Lambda
}).$ Taking into account the above historical remarks on the early
observations of vacuum-fluctuations near the boundary of a box softened by
test functions (\ref{surface}), we expect that only degrees of freedom in
the fuzzy surface around the horizon contribute to this localization-entropy.

In order to overcome the computational problems one could try to employ
similar definitions of localization-entropy which have a similar intuitive
content and avoid the direct construction of $N_{\Lambda }$. The definition
which seems to be most suitable for computations is\footnote{%
The suggestion to use this (or another closely related) definition I owe to
Heide Narnhofer who was the first to study the issue of localization-entropy 
\cite{Narn}.} that of the mathematician Kosaki who extended Araki's
definition of relative entropy\footnote{%
This entropy concept was recently successfully used by R. Longo \cite{Lo}\
in order to generalize some aspects of the Kac-Wakimoto formula from the
special setting of rational conformal theories to the theory of
superselection sectors.} by a variational formula. Araki's definition uses
his relative modular theory with respect to a von Neumann algebra $\mathcal{M%
}$%
\begin{equation}
S(\omega _{1}|\omega _{2})_{M}=-\left\langle log\Delta _{w_{1},\omega
_{2}}\right\rangle
\end{equation}
and Kosaki \cite{Kosaki} converted this (in the most general setting) into a
variational formula 
\begin{eqnarray}
S(\omega _{1}|\omega _{2})_{\mathcal{M}} &=&sup\int_{0}^{1}\left[ \frac{%
\omega (1)}{1+t}-\omega _{1}(y^{\ast }(t)y(t))-\frac{1}{t}\omega
_{2}(x^{\ast }(t)x(t))\frac{dt}{t}\right] \\
x(t) &=&1-y(t),\,\,x(t)\in \mathcal{M}  \notag
\end{eqnarray}
where in our case $\omega _{1}=\omega \times \omega ,\,\,\omega _{2}=\omega
,\,\,\mathcal{M}=\mathcal{A}\vee \mathcal{B}^{\prime }.$ An additional
simplification should be gained by studying these localization entropies
first in conformal QFT; the reason being that the modular aspects tend to be
more geometrical. They offer the additional advantage of reducing the
nuclearity (and hence the split-) property to the tracial condition 
\begin{eqnarray}
&&tre^{-\beta n^{\mu }L_{\mu }}<\infty \\
&&L_{\mu }=P_{\mu }+IP_{\mu }I
\end{eqnarray}
where $I$ denotes the geometric conformal inversion and $L_{\mu }$ turns out
to be an operator with discrete spectrum ($L_{\pm }$ are the well-known
rotation generators of the d=1+1 chiral decomposition) with $L_{0}$ positive
definite.

Let us first look at chiral conformal nets indexed by intervals on a light
ray. The simplest split is obtained by choosing an interval of length 2a
symmetrically around the origin and a slightly bigger one of length 2b
enclosing the first such that the collar size of the split situation is $%
d=b-a$ and $\mathcal{A}=\mathcal{A}(I_{a}),\,\mathcal{B}=\mathcal{A}(I_{b})$%
. It is easy to see that the localization-entropy (with any of the possible
definitions) for this situation can only depend on the harmonic ratio of
these 4 points. The modular group $\sigma _{\omega \times \omega }(t)$ is
the tensor product of the $\sigma _{\omega }^{\prime }s$ and therefore known
since the modular group for the vacuum restricted to $\mathcal{A}$ or $%
\mathcal{B}^{\prime }$ is geometric.

The nongeometric culprit is the vacuum restricted to the 2-interval algebra $%
\mathcal{A\vee B}^{\prime }.$ The ``geometrically natural'' state for $%
\mathcal{A\vee B}^{\prime }$ is not the vacuum but rather that state which
is left invariant under the diffeomorphism which leaves precisely the 4
-endpoints fixed. This is not a Moebius transformation, but it is closely
related. It is well-known that by the successive application \cite{S-W1} 
\begin{equation}
M\ddot{o}b_{2}\equiv (z\rightarrow \sqrt{z})\cdot M\ddot{o}b\cdot
(z\rightarrow z^{2})
\end{equation}
where we have used the compact $z=e^{i\varphi }$ coordinates instead if the
light ray line, one obtains a well-defined diffeomorphism (2$^{nd}$
quasisymmetric deformation of $M\ddot{o}b)$ on the circle (not in the
complex plain!). These are precisely the diffeomorphisms mentioned before in
connection with enlarging the realm of geometric modular groups beyond those
which are visible through the vacuum properties. In fact one easily check
that e.g. $U(Dil_{2}(\tau ))$ which fixes the 4 endpoints $0,1,-\infty ,-1$
and acts geometrically on chiral fields $A(x)$ (for simplicity take free
fields) leads to a limit 
\begin{eqnarray}
&&lim_{t\rightarrow \infty }\left\langle \Omega \left| A(x_{1},\tau
)...A(x_{n},\tau )\right| \Omega \right\rangle \equiv \omega
_{2}(A(x_{1})...A(x_{n})) \\
&&A(x,\tau )\equiv AdU(Dil_{2}(\tau ))A(x)  \notag
\end{eqnarray}
which defines a state $\omega _{2}$ such that ($\mathcal{A\vee B}^{\prime
},\Omega _{2})$ turns out to have $AdU(Dil_{2}(\tau ))$ as its modular
group. This state agrees precisely with the one constructed in \cite{S-W1}.
The modular groups of higher dimensional double cone in conformal theories
are known and their proximity to the two-dimensional case $(a\rightarrow
r,\,x_{\pm }\rightarrow r_{\pm }=t^{0}\pm r)$ suggest that all the modular
constructions have a higher dimensional generalization.

The calculational idea is now to compute first 
\begin{equation}
S(\omega \times \omega |\omega _{2})_{\mathcal{A\vee B}^{\prime }}
\end{equation}
and then to use the dominance of $\omega $ by $\omega _{2}$ to bound the
original split entropy. Our conjecture is that for d\TEXTsymbol{>}2 the
split entropy behaves as 
\begin{eqnarray}
S(\omega \times \omega |\omega )_{\mathcal{A\vee B}^{\prime }} &\simeq
&\left( \frac{a}{d}\right) ^{d-2} \\
\frac{a}{d} &\gg &1
\end{eqnarray}
for small d or large a such that the ratio becomes large. This would entail
the area law of the localization-entropy (associated with the causality
horizon) in conformal field theories. Since massive theories according to
common wisdom are short-distance dominated by conformal theories, the short
distance behavior in the size of the fluctuation collar d$\rightarrow 0$ has
the same divergence, and barring the presence of a competing (pathological) $%
\frac{m}{d}$ singularity, the short distance divergence remains coupled to
the surface dependence.

The main reason for emphasizing this conjecture (analogies are not yet
proofs) on the quantum version associated with the classical Bekenstein area
law\footnote{%
The causal horizons in Minkowski QFT and the Unruh effect is analogous but
not identical to black hole physics. Unruh states and Hartle-Hawking states
are different but share the thermal aspects \cite{Wald}.} in an essay like
this is that there has been hardly any subject in the last decade which has
received such an amazing amount of speculative attention going as far as
postulating some new degrees of freedom. This is quite surprising in view of
the fact that the localization-temperature has a rather mundane explanation
in terms of the KMS properties of the restricted vacuum on conventional
degrees of freedom. The situation resembles that of the speculative ideas of
how to get rid of the ultraviolet divergencies before renormalization.
Although I do not know the result in the present case, I would favor the LQP
spirit of limiting all revolutionary ideas to physical and mathematical
concepts and not to muddle with physical principles (as it was done without
success with QFT in pre-renormalization times).

\subsection{The LQP Paradigm: Quantum Measurement}

Despite its conservative way of dealing with physical principles AQFT leads
to radical change of paradigm. This is nowhere more visible than in its
relation to quantum mechanics and the measurement process. As we have seen,
the standard concepts about purity and entanglement of states \ loose their
meaning i.e. LQP is quite remote from what is done in quantum information
theory (note that the word ``local'' there has a very different meaning).
Instead of tensor factorization associated with the inside/outside
localization in quantum mechanics, the sharp relativistic boxes (double
cones) do not have pure states and an attempt to use them together with
their causally disjoint outside for the introduction of the entanglement
concept along this inside/outside division will fail: \textit{all states are
intrinsically entangled vector states thus rendering the distinction
meaningless} \cite{C-H}. Even if we use the factorization along fuzzy boxes
and their outside, we only recover these concepts at the expense of a
thermally parametrized highly mixed vacuum including all its local
excitations which constitute the natural set of states in particle physics.
As a result, most of the famous Gedankenexperiments as e.g. the
``Schr\"{o}dinger cat'' receive important qualitative modifications. But all
effects are of the ridiculously small order of the Unruh temperature (at
feasible acceleration values). Thus quite different from the recently
measured decoherence times for ``small Schr\"{o}dinger cats'' (a very small
number of photons in a cavity probed with atoms), the additional effects of
modular localization i.e. the difference between sharp and fuzzy boxes and
the entangled nature of the vacuum state with respect to any of them will
never be directly accessible. Rather one is limited to study the indirect
manifestations of e.g. the Unruh (wedge) thermality within particle physics.
In the previous section we learned that the crossing symmetry is equivalent
to the KMS thermal properties of the Hawking-Unruh effect. As such it is a
very large effect. Crossing symmetry is a property which was used in
disperion theory and the Kramer-Kronig dispersion relations for particles
were experimentally tested a long time ago.\ 

\section{A Peek at 21$^{st}$ Century Local Quantum Physics}

A glance at the future consist mostly of personal expectations and, if one
looks at the many attempts at predictions about the future and the many
resulting unfulfilled promises during the last two decades, on gets a little
bit discouraged. But just in order to prove that the modular framework is
also capable to lead to interesting conjectures and expectations let me
present some of them.

\subsection{Extension of Renormalized Perturbation Theory?}

There is certainly general agreement that gauge theories belong to the most
important contributions to 20$^{th}$ century particle physics. But on the
other hand they hardly constitute a closed mature chapter in particle
physics. In fact it is very indicative that all the important observations
about them have been made within the first 5 years after their
(re-)discovery and adaptation to the purposes of particle physics at the end
of the 60$^{ies}$ and that the rate of progress levelled off steeply
afterwards$.$ So it is natural to ask if one could expect the modular
localization method to contribute to their future development. I believe
that this question will have a positive answer.

In order to explain my reasons I find it convenient to place the problem
behind gauge theory into the slightly physically more general context of
search for renormalizable theories in which massive higher spin particles
participate. It is well known that within the causal perturbative approach
(as with any alternative approach based on Lagrangian quantization) massive
theories with spin $s\geq 1$ necessarily produce interaction densities $W$
(i.e. scalar Wick-polynomials in free fields) of at least third degree whose
operator short distance dimension $dimW\geq 5$ surpasses the value $4$
allowed by renormalizable power counting. The reason is of course that the
operator dimension of physical quantum vector fields $dimA_{\mu }=2$ is too
high as compared with its classical counterpart $dimA_{\mu }^{class}=1.$ In
fact this is not a consequence of a bad selection of a covariant field
associated with the (m,s) Wigner particle description; any other choice
would have given at least a value 2. Can one think of an had hoc
covariantization which reduces this value to 1 and at the same time does not
destroy the hope that the resulting violation of the quantum aspects of the
covariant description the spin1 Wigner particle has permanently wrecked the
physical aspects? To be more specific. is it conceivable that the ``ghost
degrees of freedom'' which achieve such a reduction of the covariantized
propagation degree act like a mysterious kind of catalyzer which are not
visible in the original problem and leave no traces in the final physical
answer but nevertheless play a beneficial intermediate role?

Everybody knows that the answer is positive and that this is formally done
with BRS ghosts in Fock space. The reason why this mathematical trick
preempts the final return to physics is the fact that it amounts to a
cohomological representation. In fact, and this is our new addition \cite
{D-S}, in the massive case this can already be implemented on the level of
the (m,s) one-particle Wigner space 
\begin{equation}
H_{Wig}=\frac{ker\frak{s}}{im\frak{s}}
\end{equation}
where $\frak{s}$ is a cohomological operator $\frak{s}^{2}=0$ which acts on
the ghost-extended Wigner space $H_{Wig}$ (not to be confused with the
pre-Tomita operator\footnote{%
This should be viewed as an operator version of the Faddeev-Popov trick.
\par
{}}$.$ The Fock space operator version of this cohomological Wigner space
representation for the operator algebras 
\begin{equation}
\mathcal{A}_{phys}=\frac{kerQ}{imQ}
\end{equation}
(where the formal operator $Q$ acts on the extended algebra by commutation)
of is nothing but a special version of the BRST formalism in which the
position of the physical space with respect to the ghost-extended space does
not change with the perturbative order. This simple formalism would not have
been available with vanishing mass because in that case the free fields in
zero order would not have been interpretable as the in-fields in the sense
of time-dependent scattering theory (appearance of infrared-divergencies).
Massive field theories, even if analytically more complicated, are
conceptually simpler. The findings of this way of looking at spin=1
interactions can be described as follows \cite{D-S}

\begin{itemize}
\item  Physical consistency within the renormalizability requirement demands
the existence of additional physical degrees of freedom which in their
simplest (and probably only) realizations are scalar particles as in the
Higgs mechanism of gauge theories but without vacuum condensates which was
characteristic of that mechanism. The intrinsic role of this field is the
implementation of the Schwinger-Swieca charge screening.

\item  Some of the ``elementary'' physical fields (i.e. those which
interpolate the perturbative particles) appear composite in the extended
formalism. The rules for a direct characterization of physical fields remain
presently complicated and their intrinsic nature is essentially not
understood; they certainly do not follow simple invariance rules as the
fixpoint algebras under a group symmetry, rather their representation in the
extended formalism lead to ever changing linear combinations of composites.

\item  Apart from the renormalization induced self-interaction of the scalar
Higgs analogues, the renormalization requirement is more restrictive%
\footnote{%
Classically the appearance of more Lorentz indices for increasing spin would
enlarge the possibilities of invariant couplings} than expected and governed
by just one coupling strength. In this sense the renormalization within the
causal setting leads to gauge structure of the coupling: the gauge groups
are not put in but result from the assumed particle multiplicities in
conjunction with the cohomological trick which is part of renormalization
and has nothing to do with group symmetry. In the standard presentation this
appears the other way around and goes with the dictum: local gauge symmetry
implies renormalizability.
\end{itemize}

Here we have tacitly assumed that there are several mutually interacting
spin one objects in order to avoid the abelian case. In the case of abelian
vectormesons there are two renormalizable models: the above one in which all
physical matter fields (including the new scalar degree of freedom) have
their expected short-distance dimension, and ``massive QED'' for which e.g.
the physical spinor matter field has an ever increasing short distance
behavior (i.e. it is an unrenormalizable field within a otherwise
renormalizable theory) or a renormalizable representation in its unphysical
(``gauge dependent'') extended realization.

This last remark suggests the following question: \textit{is it conceivable
that there are theories which are partially renormalizable i.e. in which
suitably restricted observable subalgebras have a normal short-distance
behavior?} Could it be that Lagrangian field coordinates (in particular if
they belong to higher spin) are not minimal in the sense of short distances
i.e. the same theory allows better behaved field coordinatizations which are
not Lagrangian? What at all is the physical meaning of ``short distance'' in
a field-coordinatization-free formulation in the LQP spirit; short distance
behavior of what?

Especially this last question brings us back to the main theme of this
essay: the modular localization approach. Since the wedge-localized algebra
is a field-coordinate independent object and the local net is obtained by
intersection of algebras, such a procedure would directly confront these
questions. There is no worry about ghosts reappearing in such a setting
since the short distance behavior of pointlike objects has gone which was
their reason d'etre.

\textit{In fact the modular formalism can be interpreted as an extension of
the Wigner theory to the realm of interactions}. Its starting point, the
wedge algebra is on-shell\footnote{%
The fact that on-shell quantities are free of ghosts has been used in the
tree approximation unitarity S-matrix arguments in favour of a gauge
theoretic description of vectormesons.} The improvement of localization i.e.
the transition to off-shell double cone algebras is done by intersections
and in no way calls for ghosts or touches in any other way the standard
short-distance issue. So the interesting remaining problem is: can these
ideas be supplemented with some new perturbative technology which extends
the realm of the standard renormalization theory. This implies in particular
the reproduction of the correct old results.

Looking back to the particle physics of the 60$^{ies},$ one even gets the
impression that the ill-fated S-matrix bootstrap approach was an attempt in
this direction. For an outside observer as the present author it is too hard
to find out why the program of perturbative constructions of crossing
symmetric S-matrices by pure on-shell methods failed. If the reason was the
lack of additional concepts which are capable to converts the loose ideas
about crossing symmetry and its analyticity requirements together with
unitarity into an efficient formalism, then the new modular framework should
do much better. Indeed the transition from crossing symmetry to the thermal
KMS properties for the correlations of PFG's as in section 2 is expected to
give a physically richer and formally more systematic starting point than
the old bootstrap approach to particle physics. Needless to say that the
modular approach does not support the ``cleansing ideology'' of the S-matrix
bootstrap approach against off-shell concepts from QFT. To the contrary, the
modular structure, more than any other method of particle physics, places
causality and spacetime localization back onto the centre of the stage. In
doing this it sheds new and quite unexpected light on the old
on-shell/off-shell dichotomy of particle physics which remained unaccessible
to differential geometric methods. It elevates the intrinsic spirit of
Wigner's 1939 quantum theory of free relativistic particles to the level of
interacting local quantum physics.

It is well-known that infrared problems indicate a change of the Wigner
particle picture \cite{Bu-Por}. In the present proposal this shows up in the
appearance of violent (off-shell) infrared divergencies due to the breakdown
of the Fock-space structure and the loss of physically defined (by
scattering theory) reference (free) fields. In terms of the above BRST-like
cohomological extension in the setting of point-like fields this means that
e.g. the physical $\psi $- fields (describing the spinor matter) which are
equal to the original $\psi $-fields, do not have zero mass limits. This is
a manifestation of of charge liberation which is the inverse mechanism to
the afore-mentioned Schwinger-Swieca charge screening. From general LQP
structure results we expect that charge-carrying fields in QED-like theories
do not admit compact localization since the accompanying photon clouds are
necessarily semiinfinite noncompact objects\footnote{%
The photon clouds require semiinfinite spacelike cone regions for their
localization. This is preempted on a formal level by the spacelike
Mandelstam strings of gauge theory.}. Therefore one must modify the physical 
$\psi $-fields before taking the massless limit in such a way that the
worsening of localization is preempted. It is interesting to note that this
must go together with the expected de-coupling of the Higgs-like degrees of
freedom. Both phenomena should show up after projection to the physical
perturbation theory. The infrared issue and the resulting modification of
particle structure can also be dealt with in the standard gauge approach by
seperating the algebraic aspects from those due to states \cite{DF}. Finally
one should also mention that there are other less conservative ideas which
promise to adjust the (semi)classical gauge idea directly to the
noncommutative setting. Their motivation is different from the above
attempts of extending renormalized perturbation theory beyond its present
borders (and keeping the existing renormalized results unchanged).

\subsection{Conformal Scanning?}

For the analysis of nonperturbative aspects modular theory offers a
different method which was already alluded to before, namely the reduction
of a complicated higher dimensional massive theory to a finite number of
copies of a simpler chiral conformal theory which reside in a common Hilbert
space and have a carefully tuned relative position to each other. This use
of chiral ``holography'' or ``scanning'' for general QFT is possible because
the LQP version of chiral conformal theory is more general than the standard
framework which ties chiral theories to the representation theory of a
two-dimensional energy-momentum tensor with zero physical mass. As we have
seen in section 2 the wedge algebra of a higher dimensional theory with its
light ray translations and the Lorentz-boost is naturally encoded into the
half light ray algebra. By its construction via modular inclusion the light
ray theory has automatically a conformal rotation i.e. is fully M\"{o}%
bius-covariant, i.e. the more general version leads to the same vacuum
structure as the standard.. The spectrum of the light ray translation is
gapless as it should be in a chiral conformal theory, since light cone
momenta are always gapless. The abstract chiral light ray theory does
however not possess an energy-momentum tensor with a $L_{n}$ Virasoro
algebra structure which is the hall-mark of an autonomous two-dimensional
conformal field theory. The physical mass-gap spectrum can be recovered in
the chiral light-ray holography of the wedge by a careful re-interpretation
of the geometric transformations in the wedge. In this way the light ray
translation on the lower wedge horizon becomes a ``hidden symmetry'', namely
a totally nonlocal (``fuzzy'') transformation; whereas the transversal
translations generated by $\vec{P}_{\perp }$ are presenting themselves in
the light ray world as a kind of noncompact inner symmetry. The local
generator $P_{+}$ of the light ray translation together with its hidden
counterpart $P_{-}$ and the fake internal symmetry generator $\vec{P}_{\perp
}$ define the massive physical spectrum of 
\begin{equation}
P^{\mu }P_{\mu }\equiv P_{+}P_{-}-\vec{P}_{\perp }^{2}
\end{equation}
In view of this additional partially hidden structure of chiral theories
originating from holographic projections of higher dimensional massive ones
as compared to the standard ones (based on the existence of a Virasoro type
energy-momentum tensor), it is sometimes helpful to picture the chiral
projections as associated with the d-1 dimensional (upper) horizon of the
wedge. This does no harm as long as one remains aware of the fact that this
picture does not include the net structure associated with the $P_{-}$ and $%
\vec{P}_{\perp }$ translations. The remaining L-transformations which are
not symmetries of the standard wedge W, are transforming $\mathcal{A}(W)$
into a differently positioned $\mathcal{A}(W^{\prime })$ i.e. are
isomorphisms within the total algebra $B(H).\,\,$For d=2+1 one only needs
one particular operator from the one-parametric family of ``tilting'' boosts
which fix the upper light ray. Such transformations are well-known from the
Wigner ``little group'' of light like vectors. In the present case of d=2+1
the little group is generated by just one ``translation'' (within the
L-group). Any special transformation from this 1-parametric family different
from the identity will via a $W^{\prime }$ and its holographic light cone
projection $\mathcal{A}^{\prime }(R_{+})$ lead to an isomorphism within $B(H)
$ of $\mathcal{A}(R_{+})$ to $\mathcal{A}^{\prime }(R_{+}).$ It is plausible
that such isomorphism between two differently positioned light ray algebras
can encode the missing covariances and net structure. This can be
demonstrated by applying the theory of modular intersections to the two
light ray (alias wedge) algebras. In dimension d one needs precisely d-2
specially positioned chiral theories in order to recover the full Poincar%
\'{e} symmetry and the d-dimensional net structure. As far as counting
parameters is concerned, this corresponds precisely to a light front
holography onto the horizon of the wedge, but a better picture is that of a
scanning by d-1 (isomorphic) copies of a chiral theory. In order to apply
these ideas for practical constructive purposes in higher dimensional field
theories, one should look for an extension of the notion of modular
intersection to more than two algebras. Using a similar historical analogy
as above (modular wedge localization method $\simeq $ extension of Wigner
representation theory), it is tempting to interpret the modular holography
as a clarification and extension of light cone (or p$\rightarrow \infty $
frame) physics.

In order to accomplish such a program, the understanding of chiral conformal
field theories themselves should be improved. Its present sectarian role
with respect to higher dimensional QFT and the general principles is clearly
caused by the heavy reliance on special algebras (energy-momentum tensor,
affine, current) which have no higher dimensional counterpart. On the other
hand the theory of superselection rules and their consequences for
particle/field statistics is common to all theories. In the particular case
at hand \cite{FRS} the admissible statistics belongs to the braid group and
can be in fact classified by Markov traces on the braid group which via GNS
construction lead to combinatorical type II von Neumann algebras (sometimes
inappropriately called  ``topological field theories''). They contain the
statistics information in such a way that the permutation group statistics
emerges as a special case. The missing field theoretic part is the use of
the quantized statistics (the statistical dimensions follow the famous
trigonometric Jones formula) for the construction of the spacetime carriers
of these superselected charges. The ultimate test should consist in the
derivation of FQS-quantization of the central charge from the physically
more universal statistics quantization. It is clear that the modular theory
must play an important role \cite{S3}.

\subsection{A higher dimensional Theory of Anomalous Dimension?}

In order to avoid the impression that the conservative attitude of LQP with
respect to physical principles prevents addressing presently fashionable
subjects, I would like to explain some speculative ideas on so-called SYM
models. This is clearly part of the general question of nontrivial aspects
of higher dimensional conformal QFT. As in the well-studied d=1+1 conformal
theories, interpolating local fields which create Wigner particles are
necessarily canonical free fields. Hence nontrivial fields cannot be
associated with Wigner particles and must have noncanonical anomalous
dimensions (which at best can be associated with infraparticles). So the
first step in unraveling the structure of d\TEXTsymbol{>}1+1 conformal
theories should be the understanding of its spectrum of anomalous
dimensions. For d=1+1 conformal models such a theory of anomalous dimension
(critical indices of associated critical statistical mechanics) exists;
these numbers are determined (modulo $2\pi )$ by the statistical phases of
the braid group statistics of the fields (the R-matrices of the exchange
algebra). The classification of physically admissable braid group statistics
is a well-defined mathematical problem which can be separated from the
spacetime aspects of QFT and treated by the technique of Markov-traces. The
construction of nets fulfilling exchange algebra relations can be converted
into a well-defined problem of modular theory. Can one achieve a similar
situation with respect to anomalous dimensions ($\simeq $ critical indices)
in higher dimensional conformal theories? The answer is positive for
theories which admit observable algebras which fulfil timelike commutativity
i.e. which propagate only in lightlike directions (Huygens principle) as
zero mass free fields. There are arguments that by choosing the observable
algebra sufficiently small, this can always be achieved. One would like to
interpret anomalous dimension fields as carriers of superselection charges
associates with the timelike local observable algebra and one glance at the
two-point function reveals that one should expect timelike braidgroup
commutation relations associated with the timelike ordering structure%
\footnote{%
In fact the time-like net in the forward light cone admits a projection onto
the timelike line which is a chiral conformal theory without the Virasoro
structure.}. This is indeed what a sytematic DHR analysis in terms of
localized endomorphisms confirms. We obtain the whole superselection
formalism with braidgroup (R-matrix) commutation relations except that the
statistic interpretation is missing: from the viewpoint of spacelike
commutation relations we are dealing with Bosons/Fermions. The two- and
three-point functions of the observable fields suffer the usual conformal
restrictions i.e. they are determined by their dimensions modulo a
normalization constant which carries the memory about the interaction. If
supersymmetry ``protects'' these parameters against changes due to
interactions, then such a model is in dangerous proximity of a free field
theory.

My conviction that the present modular framework and more generally the LQP
approach will have a rich future stems primarily from the fact that the
intrinsic logic of LQP is strong and convincing that it appears a safer
guide than that obtained from the quantization approach. Whereas the
canonical formalism, the interaction picture, the formalism of time-ordering
etc. can (and has been) be used outside of relativistic QFT, the modular
approach is \textit{totally specific for real time LQP}. In fact it is the
only truly noncommutative entrance into QFT which came really from physics
(rather than physical illustrations of mathematical concepts as done e.g.
with noncommutative geometry). Admittedly, it is an area, which because of
its strong conceptual roots and demanding mathematical apparatus is not easy
to enter; neither does the subject render itself to fast publications. But
in compensation, even if progress at times is very slow, it carries a
conceptual profoundness and mathematical solidity which, if coupled with the
belief in the guiding power of physical principles (especially through times
of crisis), is hard to match.

A superficial observer would conclude from the present account that particle
physics is strong and healthy with a promising 21$^{st}$ century future.
Such an observer has missed to notice the radical change of values which
also profoundly altered the exact sciences. An outburst of stunning
creativity as it happened at the beginning of last century (Plank, Einstein,
Bohr, Heisenberg) is only possible under very special sociological
conditions in which the search for scientific truth and universality has a
high social ranking and were new emerging ideas in sciences were always
confronted with historical and traditional aspects. These are not
necessarily the good times as the explosion of sciences and the arts in the
imperial as well as in the humiliated post-war Germany of theWeimar republic
shows. \ \ 

Present sociology and philosophy of life is totally different. The high
social ranking of shareholder-values and globalization over productive
values in modern capitalism has found its counterpart also in particle
physics. It consists of using ones knowledge, including mathematical
sophistication primarily for improving ones status within a scientific
community and not for the benefit of furthering science. This works because
it is tacitely accepted by a majority. In earlier times there still existed
a perceived difference of ``physics'' and what at one or the other time
``physicist were doing'', whereas nowadays this distinction disappeared. How
can one otherwise explain that theories which already exist for 30 years and
besides making their inventors famous never contributed anything to particle
physics enjoy such popularity? And how can one explain that rewards are
starting to be given to inventors, thus setting examples for the young
generation? The acquired profound knowledge about quantum field theory is
now rapidly getting lost and it is a truly amazing experience to meet young
people who do not have the slightest idea about scattering theory,
dispersion theory and the Wigner particle theory although they know more
than necessary about Calabi-Yao spaces, Riemann surfaces and all those
theories which hide behind big Latin Letters. At most places it is already
impossible to have a carrier in physics outside these trends; the academic
freedom is rapidly loosing its economic basis. Fast returns as with
shareholder values are incompatible with the flourishing of particle
physics. If this trend continues another 10 years, the profound knowledge
about real problems of 20$^{th}$ century particle physics and QFT will have
been lost with the young generation. Even if one believes that truths in
exact sciences will always eventually find its way, one does not want to be
proven correct on such a rather pessimistic outlook.

Acknowledgements:

A good part of the actualization of an older 1998 manuscript was done at the
ESI following an invitation of J. Yngvason. I thank Prof. Yngvason for this
invitation and Profs. Thirring and Narnhofer for several discussions.


\begin{thebibliography}{99}
\bibitem{Haag}  R. Haag, ``Local Quantum Physics'', Springer Verlag (1992)

\bibitem{Weinberg}  S. Weinberg, ``What is Quantum Field Theory , and What
Did We Think It Is?'', hep-th/9702027

\bibitem{Wei}  S. Weinberg, \textit{The Quantum Theory of Fields, I,}
Cambridge University Press 1995

\bibitem{B-H}  D. Buchholz and R. Haag, \textit{The Quest for Understanding
in Relativistic Quantum Physics}, University of Goettingen preprint October
1999, Invited Contribution to the Issue 2000 of JMP

\bibitem{Bor}  H.-J. Borchers, \textit{On the Revolutionization of Quantum
Field Theory by Tomitas Modular Theory}, University of Goettingen preprint,
April 1999, Invited Contribution to the Issue 2000 of JMP

\bibitem{Sch1}  B. Schroer, Annals of Physics \textbf{275}, No2 (1999) 190
and previous paper of the same author quoted therein 

B. Schroer, ``New Concepts in Particle Physics from Solution of an Old
Problem'', hep-th/9908021

\bibitem{BGL}  R. Brunetti, D. Guido and R. Longo, ''First quantization via
BW property'', in preparation

\bibitem{S-W2}  B. Schroer and H.-W. Wiesbrock, RMP Vol 12 No 2 (Feb 2000)
301-326

\bibitem{Za}  A. B. Zamolodchikov, Int. J. of Mod. Phys. \textbf{A1}, (1989)
1235

\bibitem{Fring}  A. Fring, Int.J.Mod.Phys. A11 (1996) 1337

\bibitem{STW}  B. Schroer, T.T. Truong and P. Weisz, Phys. Lett. \textbf{B63 
}(1976) 422

\bibitem{BKTW}  M. Karowski, H. J. Thun, T. T. Truong and P. Weisz, Phys.
Lett. \textbf{B67 }(1977) 321

\bibitem{Zam}  A. B. Zamolodchikov, JETP Lett. \textbf{25} (1977) 468

\bibitem{S-W1}  B. Schroer and H.-W. Wiesbrock, RMP Vol \textbf{12} No 1
(Jan 2000) 139

\bibitem{BFKZ}  H. Babujian, A. Fring. M. Karowski and A. Zapletal, Nucl.
Phys. B\textbf{538}, (1999) \%\S \%

\bibitem{GLRV}  D. Guido, R. Longo, J.E. Roberts and R. Verch, \textit{%
Charged Sectors, Spin and Statistics in Quantum Field Theory on Curved
Spacetimes}, math-ph/9906019, and referces therein

\bibitem{Wightman}  R.F. Streater and A.S. Wightman, \textit{PCT, Spin and
Statistics and all That}, Benjamin 1964

\bibitem{Mund}  J. Mund, Lett.Math.Phys. 43 (1998) 319-328

\bibitem{BBS}  D. Buchholz, H.-J. Borchers and B. Schroer, ``Polarization
Free Generators and the S-Matrix'', in preparation

\bibitem{S2}  B. Schroer, ''New Concepts in Particle Physics from Solution
of an Old Problem'', CBPF Rio de Janeiro preprint, October 1998

\bibitem{White}  A. R.White, ``The Past and Future of S-Matrix Theory'',
hep-ph/0002303

\bibitem{Wies}  H.-W. Wiesbrock, Comm. Math. Phys. \textbf{157 }p 83 (1993) 
\textbf{Erratum }Comm. Math. Phys. \textbf{184, }683 (1997), Lett. Math.
Phys. \textbf{39, } 203 (1997)

\bibitem{S}  B. Schroer, \ Int. J. of Mod. Phys. \textbf{B6}, (1992) 2041

\bibitem{Bo}  H. J. Borchers, Comm. Math. Phys. \textbf{143 }p. 315 (1992)

Ann. Henri Poincar\'{e} \textbf{63} p 331 (1995)

\bibitem{BDFS}  D. Buchholz, O. Dreyer, M. Florig and S. J. Summers

\bibitem{S-W3}  B. Schroer and H.-W. Wiesbrock, ``Looking beyond the Thermal
Horizon: Hidden Symmetries in Chiral Models'', to appear in RMP Vol 12 No 3
(March 2000

\bibitem{B-Y}  H.-J. Borchers, \ ``On Thermal States of (1+1)-dimensional
Quantum Systems'' ESI preprint Nr. 788, (1999)

\bibitem{Bu}  D. Buchholz and P. Junglas, Commun. Math. Phys. \textbf{121},
(1989) 255

\bibitem{BDLR}  D. Buchholz, S. Doplicher, R. Longo and J.H. Roberts, Rev.
Math. Phys. Special Issue,

\bibitem{Do-Lo}  S. Doplicher and R. Longo, Invent. math. \textbf{75},
(1984) 493

\bibitem{Connes}  A. Connes, Ann. Inst. Fourier \textbf{126}, (1974) 121

\bibitem{C-H}  Rob Clifton and Hans Halvorson, ``Entanglement and open
Systems in Algebraic Quantum Field Theory'', University of Pittsburgh
preprint Jan.2000

\bibitem{Narn}  H. Narnhofer, in ``The State of Matter'', ed. by M. Aizenman
and H. Araki (Wold-Scientific, Singapore) 1994

\bibitem{Kosaki}  H. Kosaki, J. Operator Theory \textbf{16}, (1986) 335

\bibitem{Lo}  R. Longo, ``The Bisognano-Wichmann Theorem for charged states
and the Conformal Boundary of a Black Hole'' math-ph/0001034

\bibitem{Wald}  R. M. Wald, ``Quantum Field theory in Curved Spacetime and
Black Hole Thermodynamics'', University of Chicago Press 1994

\bibitem{D-S}  M. Duetsch and B. Schroer; \textit{Massive Vectormesons and
Gauge Theory}, hep-th/9906089

\bibitem{Bu-Por}  D. Buchholz, M. Porrmann and U. Stein, Phys. Lett.B267
(1991) 377

\bibitem{DF}  M. Duetsch and K. Fredenhagen, Commun.Math.Phys. 203 (1999) 71

\bibitem{FRS}  K. Fredenhagen, K-H Rehren and B. Schroer, Rev. Math. Phys., 
\textbf{Special Issue} (1992) and references therein. 

\bibitem{S3}  B. Schroer, ``Local Quantum Theory beyond Quantization'',
hep-th/9912008
\end{thebibliography}
\end{document}